\documentclass[3p,twocolumn,preprint]{elsarticle}

\usepackage{lineno,hyperref}
\modulolinenumbers[5]

\usepackage{tikz}
\usepackage{pgfplots}
\usepackage{pgfmath}

\definecolor{RYB1}{RGB}{166,206,227}  
\definecolor{RYB2}{RGB}{31,120,180}   
\definecolor{RYB3}{RGB}{178,223,138}  
\definecolor{RYB4}{RGB}{51,160,44}    
\definecolor{RYB5}{RGB}{251,154,153}  
\definecolor{RYB6}{RGB}{227,26,28}    
\definecolor{RYB7}{RGB}{253,191,111}  
\definecolor{RYB8}{RGB}{255,127,0}    
\definecolor{RYB9}{RGB}{202,178,214}  
\definecolor{RYB10}{RGB}{160,60,140}  
\definecolor{RYB11}{RGB}{255,255,153} 
\definecolor{RYB12}{RGB}{177,89,40}   
\definecolor{mfem@blue}{RGB}{100,150,230}
\definecolor{mfem@green}{RGB}{75,200,75}
\definecolor{mfem@red}{RGB}{200,75,75}
\definecolor{mfem@orange}{RGB}{252,186,3}

\definecolor{code}{rgb}{0, 0, 0}
\newcommand{\code}[1]{\texttt{\small\color{code} #1}}

\usepackage{soul}
\definecolor{cola}{RGB}{28, 97, 166}
\definecolor{colb}{RGB}{192, 0, 3}

\newcommand{\reva}[1]{{#1}}
\newcommand{\revb}[1]{{#1}}

\pgfplotscreateplotcyclelist{will}{%
black,every mark/.append style={fill=white},mark=*\\%
RYB4!50!black,every mark/.append style={fill=RYB4},mark=*\\%
RYB6!50!black,every mark/.append style={fill=RYB6},mark=*\\%
RYB8!50!black,every mark/.append style={fill=RYB8},mark=*\\%
RYB10!50!black,every mark/.append style={fill=RYB10},mark=*\\%
RYB12!50!black,every mark/.append style={fill=RYB12},mark=*\\%
black,every mark/.append style={fill=white!50!black},mark=*\\%
RYB2!50!black,every mark/.append style={fill=RYB2},mark=*\\%
RYB4!50!black,densely dashed,every mark/.append style={fill=RYB4},mark=*\\%
RYB6!50!black,densely dashed,every mark/.append style={fill=RYB6},mark=*\\%
RYB8!50!black,densely dashed,every mark/.append style={fill=RYB8},mark=*\\%
RYB10!50!black,densely dashed,every mark/.append style={fill=RYB10},mark=*\\%
RYB12!50!black,densely dashed,every mark/.append style={fill=RYB12},mark=*\\%
black,densely dashed,every mark/.append style={fill=white!50!black},mark=*\\%
RYB2!50!black,densely dashed,every mark/.append style={fill=RYB2},mark=*\\%
}

\begin{document}

\begin{frontmatter}

\title{GPU Algorithms for Efficient Exascale Discretizations}

\author[5]{Ahmad Abdelfattah}
\author[6]{Valeria Barra}
\author[5]{Natalie Beams}
\author[15]{Ryan Bleile}
\author[6]{Jed Brown}
\author[1]{Jean-Sylvain Camier}
\author[16]{Robert Carson}
\author[9]{Noel Chalmers}
\author[1]{Veselin Dobrev}
\author[1]{Yohann Dudouit}
\author[2,3,4]{Paul Fischer}
\author[19]{Ali Karakus}
\author[2]{Stefan Kerkemeier}
\author[1]{Tzanio Kolev\corref{mycorrespondingauthor}}\ead{tzanio@llnl.gov}
\author[2]{Yu-Hsiang Lan}
\author[2,17]{Elia Merzari}
\author[2]{Misun Min}
\author[3]{Malachi Phillips}
\author[3]{Thilina Rathnayake}
\author[15]{Robert Rieben}
\author[15]{Thomas Stitt}
\author[2,18]{Ananias Tomboulides}
\author[5]{Stanimire Tomov}
\author[1]{Vladimir Tomov}
\author[15]{Arturo Vargas}
\author[7]{Tim Warburton}
\author[15]{Kenneth Weiss}

\cortext[mycorrespondingauthor]{Corresponding author}

\address[1]{Center for Applied Scientific Computing, Lawrence Livermore National Laboratory, Livermore, CA 94550}
\address[2]{Mathematics and Computer Science, Argonne National Laboratory, Lemont, IL 60439}
\address[3]{Department of Computer Science, University of Illinois at Urbana-Champaign, Urbana, IL 61801}
\address[4]{Department of Mechanical Science and Engineering, University of Illinois at Urbana-Champaign, Urbana, IL 61801}
\address[5]{Innovative Computing Laboratory, University of Tennessee, Knoxville, TN 37996}
\address[6]{Department of Computer Science, University of Colorado, Boulder, CO 80309}
\address[7]{Department of Mathematics, Virginia Tech, Blacksburg, VA 24061}
\address[9]{AMD Research, Advanced Micro Devices Inc., Austin, TX 78735}
\address[15]{Weapons and Complex Integration, Lawrence Livermore National Laboratory, Livermore, CA 94550}
\address[16]{Computational Engineering Division, Lawrence Livermore National Laboratory, Livermore, CA 94550}
\address[17]{Department of Nuclear Engineering, Penn State, PA 16802}
\address[18]{Department of Mechanical Engineering, Aristotle University of Thessaloniki, Greece 54124}
\address[19]{Mechanical Engineering Department, Middle East Technical University, Ankara, Turkey 06800}

\begin{abstract}
In this paper we describe the research and development activities in the Center
for Efficient Exascale Discretization within the US Exascale Computing Project,
targeting state-of-the-art high-order finite-element algorithms for high-order
applications on GPU-accelerated platforms. We discuss the GPU developments in
several components of the CEED software stack, including the libCEED, MAGMA,
MFEM, libParanumal, and Nek projects. We report performance and capability
improvements in several CEED-enabled applications on both NVIDIA and AMD GPU
systems.
\end{abstract}


\end{frontmatter}


\section{Introduction} \label{sec:introduction}

Exascale computing will provide scientists and engineers with an advanced tool
to explore physical phenomena over a large range of scales and in complex
domains. To maximize this potential, the simulation codes must be efficient in
their use of data movement, both in terms of implementation and algorithmic
complexity. The DOE Center for Efficient Exascale Discretizations (CEED)
\cite{ceed, ceed_ecp_special_2020} seeks to meet both of these goals by
providing highly performant libraries for high-order discretizations on
GPU-based compute nodes that form the basis for current- and next-generation
HPC platforms.

Central to CEED is the use of matrix-free high-order finite element
discretizations, which require only $O(n)$ data movement and yield
exponential convergence rates, $O(h^p$), for $p$th-order approximations to
solutions having \reva{sufficient} regularity. With the number of \reva{degrees
of freedom} scaling as \reva{$n = O\big((p/h)^{d}\big)$}, in $d$ dimensions,
convergence through increased $p$ offers clear advantages over simple
reductions in the grid spacing $h$.
\reva{
Kreiss and Oliger \cite{Kreiss72} noted early on the particular relevance
of increased approximation order in controlling cumulative dispersion errors
for large-scale transport problems where propagated feature sizes $\lambda$ are
much smaller than the domain length, $L$,  which is clearly in the scope of
problems that are enabled by exascale architectures.
Although exponential convergence is lost in many practical applications that
lack regularity, high-order methods still provide favorable error constants
with respect to norms and insidious sources of error such as numerical
dispersion, and use of $hp$ methods can sometimes restore exponential
convergence \cite{babuska1994hpfem}.
}
Efficient implementation of methods of all orders, with a particular emphasis
on high-order, is the principal objective of the CEED efforts.

\def\Oh{{\hat \Omega}}
\def\br{{\bf r}}
\def\bx{{\bf x}}
\def\uu{{\underline u}}
\def\uw{{\underline w}}
\newcommand{\pp}[2]{\frac{\partial #1}{\partial #2} }
\newcommand{\dd}[2]{\frac{d #1}{d #2} }
\def\Dh{{\hat D}}
\def\bD{{\bf D}}
\def\bG{{\bf G}}
\def\Ih{{\hat I}}

\reva{
Practical application of spectral methods for complex domains was first
considered by Orszag \cite{sao80}, who laid out several essential elements for
performant implementations. The principal feature was the use of $p$th-order
tensor-product polynomial approximations in a $d$-dimensional reference domain,
$\br \in \Oh=[-1,1]^d$, transformed to a complex domain, $\Omega$, through an
invertible map, $\bx(\br)$.
  Unlike Fourier bases, stable polynomial bases\footnote{Stable bases include
orthogonal polynomials or Lagrange polynomials based on Gauss-type quadrature
points.} can yield exponential convergence for non-periodic boundary
conditions, provided the solution has sufficient regularity \cite{dgso77}.
  Orszag \cite{sao80} noted that, although separability was lost in the
transformed domain, the forward operator evaluation could still be effected
efficiently using tensor-product sum factorization. He thus suggested using
conjugate gradient iteration, preconditioned with spectrally-equivalent
low-order (i.e., sparse) operators, to yield high-order accuracy at low-order
costs. All three of these ideas---orthogonal-polynomial based bases,
tensor-product sum factorization, and low-order preconditioners, are common
elements of modern high-order finite element codes
(e.g., \citep{ExaDG2020,Bello-Maldonado2019,gervasio10,Moxey2020,Sun2020,mfem}),
although the low-order preconditioners are commonly supplanted with
$p$-multigrid (e.g.,
\cite{KronbichlerWall2018,KronbichlerLjungkvist2019}) and/or Schwarz-overlapping
methods \cite{lottes05}.
}

Let $Q_p$ \revb{denote} the Lagrange polynomial bases on Gauss-Lobatto quadrature
points. In the case of tensor product elements, for $p^d$ degrees-of-freedom per
element, the use of tensor-product sum-factorization reduces operator evaluation
costs from $O(p^{2d})$ to (near-optimal) $O(p^{d+1})$ and memory/storage costs
from $O(p^{2d})$ to (optimal) $O(p^{d})$. Matrix-free $Q_p$ bases form the
foundation for much of the CEED software stack, including MFEM, Nek5000/RS,
libCEED, MAGMA, and libParanumal. We are also interested in high-order
discretization on (non-tensor) triangular and tetrahedral $P_p$ elements.

\revb{
Throughout the paper we use the CEED bake-off problems (BPs), introduced in
\cite{ceed_bp_paper_2020}, in order to test and compare the performance of
high-order codes. The CEED BPs are community benchmarks for matrix-free
operator evaluation of mass (BP1), stiffness (BP3) or collocated stiffness
matrix (BP5). They include a mixture of compute-intensive kernels,
nearest-neighbor communication and vector reductions that is representative of
high-order applications. See \cite{ceed_bp_paper_2020} for details.
}

Both the {\em tensor} and {\em non-tensor} cases pose unique challenges for
high-order algorithms on GPU platforms. In the following Section 2 we describe
the GPU-specific developments that are addressing these challenges in each of
the CEED open-source libraries. We follow this by a discussion in Section 3,
results from several CEED-enabled applications in Section 4, and conclusions in
Section 5.

\section{GPU Developments in the Center for Efficient Exascale Discretizations} \label{sec:ceedgpu}

The major {\em work intensive} operations for $p$th order elements are local
interpolation and differentiation, which involve tensor contractions for $Q_p$
and DGEMMs for $P_p$ elements. For example, for tensor elements in 3D derivatives at nodal points
$(r_i,s_j,t_k)\in \Oh$ take the form $u^e_r(r_i,s_j,t_k) = \sum_m \Dh_{im}
u^e_{mjk} = D_r \uu^e$, where the $u^e_{ijk}$ are the local basis coefficients
for $u^e(\br)$ on $\Omega^e$ and $\Dh$ is a dense $(p+1)\times(p+1)$ derivative
matrix.

On a CPU, performance in the {\em tensor} $Q_p$ case relies primarily on casting
the tensor contractions, which comprise 90\% of the flops, as optimized
matrix-matrix products.
On GPUs, vectorization must be expressed over the entire set of $E$ elements
local to a given GPU in order to leverage the node-parallel architecture and
amortize kernel launch overhead. Kernels thus tend to be expressed at the
operator level, such as the discrete Laplacian, $\uw_L=A_L \uu_L$, where
$A_L$=block-diag($A^e$), with $A^e=\bD^T \bG^e \bD$ the matrix-free form of the
stiffness matrix.
For deformed elements $A^e$ is dense, with $(p+1)^{6}$ entries. The factored
form, however, involves only the tensor-product derivatives, $D_r$, $D_s$, and
$D_t$, and six diagonal matrices, $G^e_{rr},$ $G^e_{rs},\; \dots\;, G^e_{tt}$ in
the symmetric tensor $\bG^e$. Including $\uu^e$, the number of reads is thus
$7(p+1)^3$ per element, with a corresponding flop count of $12(p+1)^4 + 15
(p+1)^3$ \reva{(double the first term for non-collocated quadrature)}.
The $O(p)$ \reva{arithmetic} intensity (flop-to-byte ratio)
\reva{with low-memory data structures leads to high performance on GPU and CPU architectures that have sufficient registers/cache to exhibit memory locality through the sequence of dependent operations}.

In the rest of this section, we describe the developments in each of the CEED
packages that address the multitude of issues that arise when implementing
operations of the form $A_L \uu_L$ on GPUs. For detailed description of
these operations see \cite{ceed_bp_paper_2020}.

\subsection{libCEED}
\label{subsec:libceed}
\begin{figure}[!htb]
\centering
\includegraphics[width=1.0\linewidth]{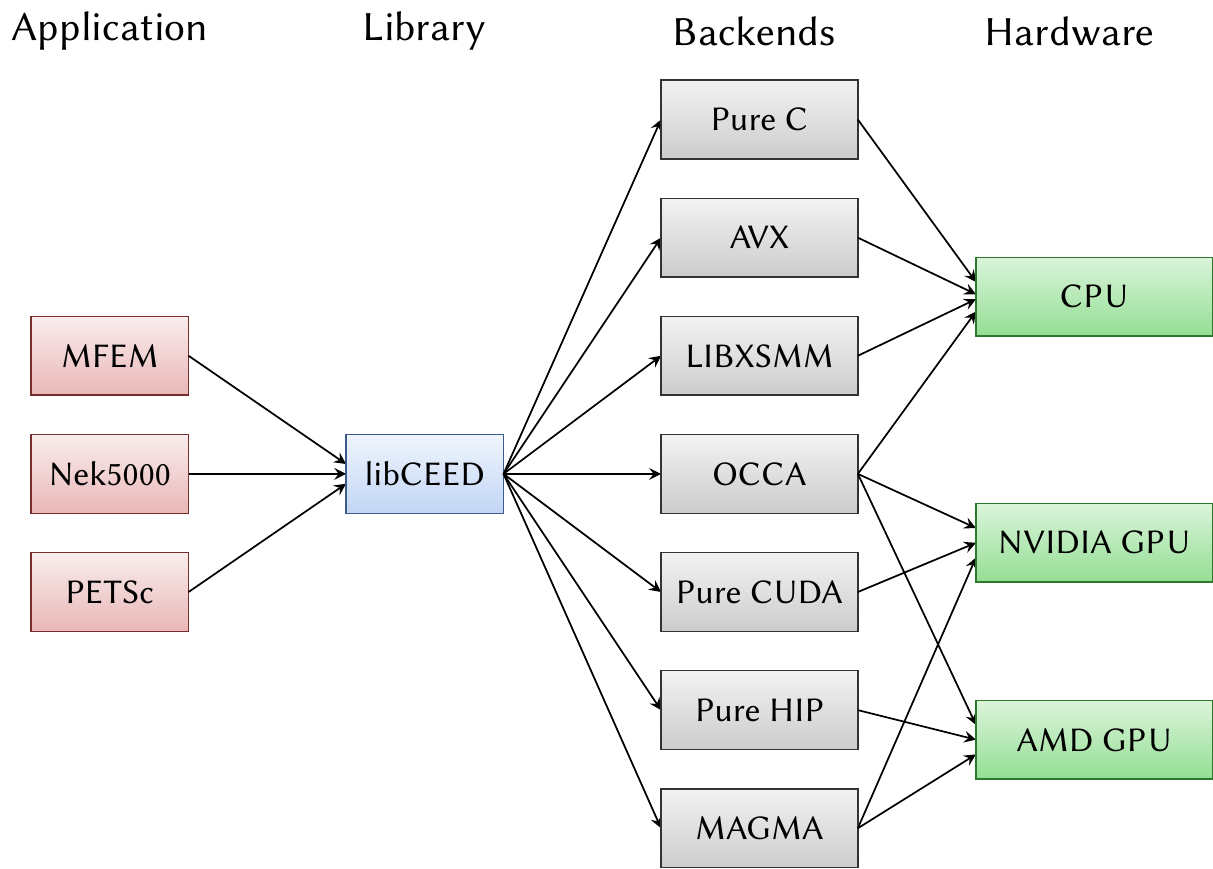}
\vspace{-6mm}
\caption{
libCEED allows different applications to share highly optimized discretization
kernels.}
\label{fig:libceed:libCEEDBackends}
\end{figure}

libCEED \cite{libceed-joss-paper, libceed-user-manual} is a new library that offers a purely algebraic
interface for matrix-free operator evaluation and supports run-time selection of
implementations tuned for a variety of computational device types, including
CPUs and GPUs. libCEED's purely algebraic interface can unobtrusively be
integrated in new and legacy software to provide performance portable
interfaces. While the focus is on high-order finite elements, the approach is
algebraic and thus applicable to other discretizations in factored
form. libCEED's role, as a lightweight portable library that allows a wide
variety of applications to share highly optimized discretization kernels, is
illustrated in Figure \ref{fig:libceed:libCEEDBackends}, where a non-exhaustive
list of specialized implementations (backends) is listed.
libCEED provides a low-level Application Programming Interface (API) for user
codes so that applications with their own discretization infrastructure (e.g.,
those in PETSc, MFEM and Nek5000) can evaluate and use the core operations
enabled by libCEED. GPU backends are available via pure CUDA or HIP
implementations, as well as the OCCA and MAGMA libraries. libCEED provides a
unified interface, so that users only need to write a single source code and can
select the desired specialized implementation at run time. Moreover, each
process or thread can instantiate an arbitrary number of backends.

Since matrix-free finite element algorithms move potentially $O(p^d)$ times less
data than algorithms using sparse matrices, where $p$ is the polynomial order,
and $d$ the dimension, $O(p^d)$ speedup can theoretically be achieved on
architectures where the matrix-free algorithms are limited in performance by the
data movements, i.e. are memory bound.

\begin{figure}[!htb]
\centering
\input{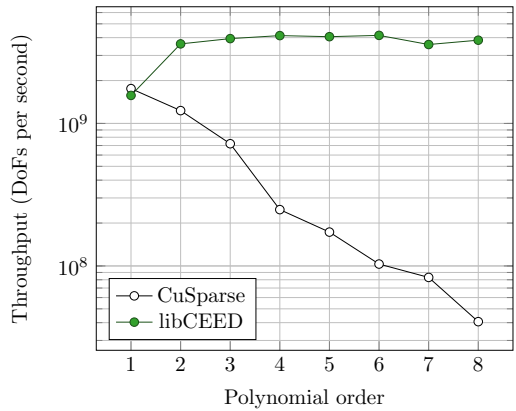}
\vspace{-6mm}
\caption{Comparison of the performance \reva{in the saturated regime} for the CEED benchmark problem BP1
\cite{ceed_bp_paper_2020} on a NVIDIA V100 using a sparse matrix (with \code{CuSparse})
against the \texttt{cuda-gen} backend of libCEED for different polynomial orders $p$.}
\label{fig:libceed:FAvsPA}
\end{figure}

The matrix-free algorithms to apply \emph{tensor} finite element operators are
completely memory bound on GPU due to their low arithmetic intensities.
$O(p^d)$ speedup can be achieved in practice for tensor finite element when
compared to a standard approach using a sparse matrix, see
Figure \ref{fig:libceed:FAvsPA}, because the GPU kernels are solely memory
bound.

However, applying \emph{non-tensor} finite element operators does not
necessarily result in memory bound GPU kernels.
\reva{The libCEED interface allows the caller to provide arrays representing the evaluation and
gradient of the basis functions along with the quadrature weights to be used.}
With the arithmetic intensity of
the matrix-free algorithms increasing in $O(p^d)$ for general non-tensor elements\revb{--compared
to $O(p)$ in the tensor element case--}GPU kernels quickly become computationally bound for high $p$ orders \revb{on sufficiently many elements},
especially in 3D. The higher number of basis functions and quadrature points for non-tensor
elements also results in operations that are more difficult to cache efficiently
on the very limited memories of GPU caches and registers. Therefore, achieving
peak performance for non-tensor finite elements is more challenging, and the
theoretical gain is not as interesting as for tensor finite elements.
\reva{We note that there exist specialized sum factorization schemes
for collapsed elements (e.g.~\cite{KarniadakisSherwin2005, Vos2010}, with recent SIMD vectorization targeting CPU performance \cite{Moxey2020}) and other fast evaluation methods such as \cite{ainsworth2011bernstein,kirby2011fast}; adding such methods to libCEED would require addition of a public interface and dedicated backend support.}

On NVIDIA GPUs the libCEED library achieves close to peak performance for
operators using tensor finite elements. The performance on the CEED benchmark
problems is comparable to state-of-the-art specialized hand tuned kernels
\cite{swirydowicz2019acceleration}.

The CUDA and HIP backends provide native support for non-tensor finite elements,
\reva{while the OCCA and MAGMA backends, which depend on their eponymous
libraries \cite{occa, magma},
also support non-tensor finite elements.  The MAGMA backend achieves the highest
performance of all libCEED GPU backends for non-tensor finite elements.}

\subsection{MAGMA}
\label{subsec:magma:nontensor}
\reva{MAGMA \cite{magma} is a high-performance linear algebra library that includes
LAPACK for GPUs, BLAS, sparse iterative solvers, and many other general matrix
computation kernels. The batched computations provided by MAGMA can be generalized to provide
highly efficient tensor computations~\cite{abdelfattah2016high}.  While the libCEED MAGMA backend
contains specialized tensor basis kernels separate from the MAGMA library itself,
the library's batched GEMM capabilities are used directly to optimize non-tensor basis computations,
with a goal of hardware portability \cite{magma_libceed2020}.}
\reva{In contrast to the recent CPU-based work of Sun et al.~\cite{Sun2020}, which applies code
generation and transformations toward SIMD vectorization across batches of (tensor and non-tensor) element
kernels, the MAGMA backend's batched computations leverage standard library
BLAS routines for the non-tensor basis actions within libCEED's algebraic framework.}

\begin{figure}[!htb]
\centering
\includegraphics[width=0.8\linewidth]{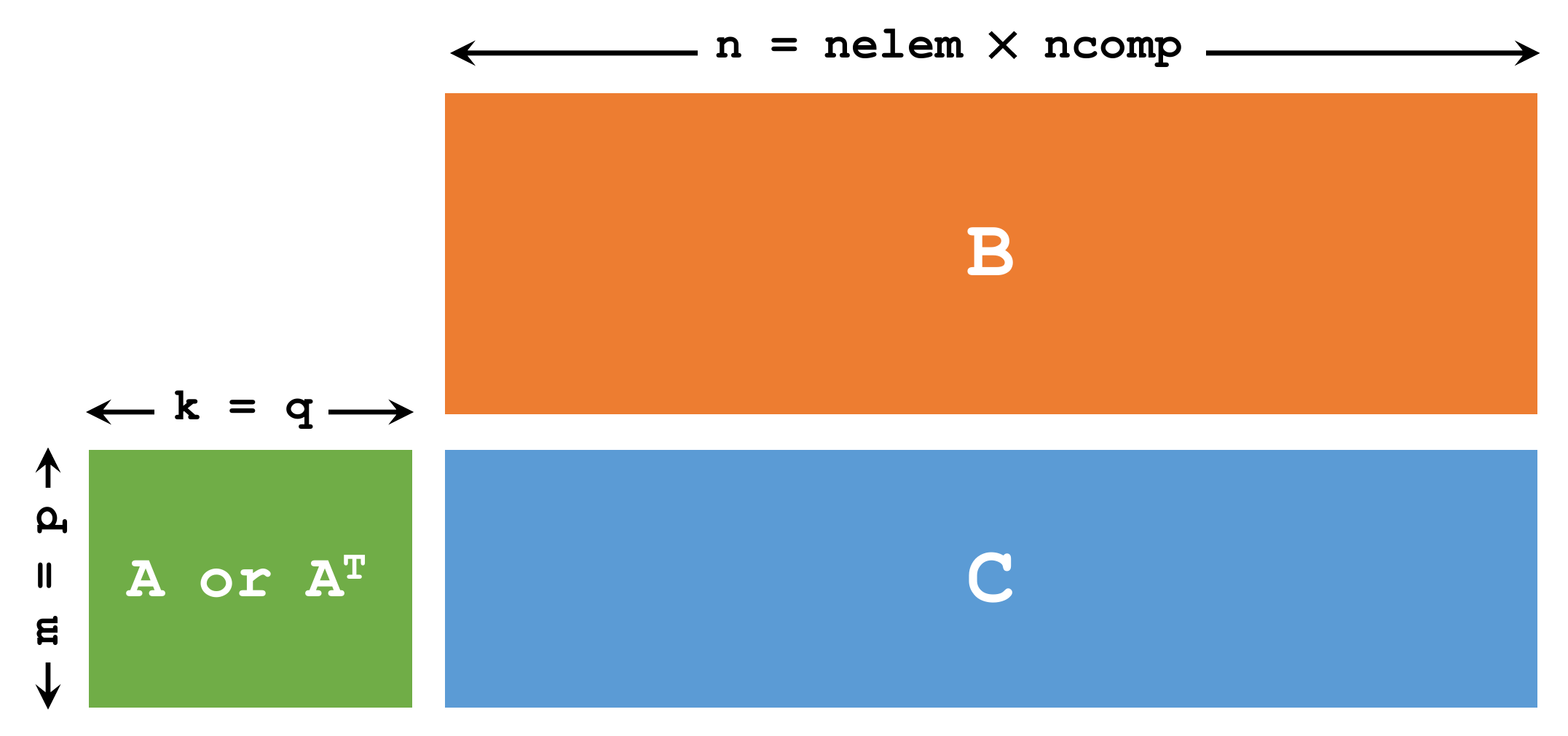}
\vspace{-2mm}
\caption{Shape of the DGEMM operation for the non-tensor basis action in
libCEED.}
\label{fig:magma:dgemm_shape}
\end{figure}

\reva{As the non-tensor basis computations in libCEED are basis- and quadrature-rule-agnostic,
the full interpolation or gradient matrices must be applied for an input vector for every
element, rather than performing a series of small tensor contractions.
If we consider all elements local to the process at the same time
({\tt nelem}), we can reshape the input vector to be a matrix of size
$d \times P \times {\tt nelem} \times {\tt ncomp}$, with each column
corresponding to one component for one element, and $P$ and $Q$ representing the total
number of basis and quadrature points in the element, respectively.} Now the application of the
interpolation or gradient matrix action for all the elements is easily
represented by one standard general matrix-matrix multiplication, $C = AB$, as
represented in Figure \ref{fig:magma:dgemm_shape} for input $B$, output $C$, and
basis matrix $A$.

Figure~\ref{fig:magma:dgemm_shape} shows a typical shape for the GEMM call in a
libCEED non-tensor basis computation, with dimensions $(m, n, k)$. Here $m$ and
$k$ are relatively small, as they are tied to the number of basis nodes or
quadrature points in one element; $n$ can potentially be orders of magnitude
larger, as it depends on the number of elements in the local operator. Since
vendor-provided GEMM routines are generally designed to achieve maximum
performance for square matrices, these unbalanced dimensions may prevent the
GEMM operation from reaching the GPU peak performance. Therefore, to make the
best possible use of the available BLAS libraries, we also consider performing
the GEMM operation in Figure~\ref{fig:magma:dgemm_shape} as a batched GEMM
operation, split across the $n$ dimension, so that
each batched operation has the same $A$ matrix, but uses submatrices $\hat{B}$
and $\hat{C}$ of $B$ and $C$, with $\eta$ columns each. The additional parameter
of batch size $\eta$ increases the space in which we can search for the best
possible parameter set, given dimensions $m$ and $k$, with the goal of creating
a more balanced workload for the GPU. Transforming the GEMM in
Figure~\ref{fig:magma:dgemm_shape} into a batched GEMM does not require setting
up pointer arrays that may impact the performance, and thus does not add any
overheads. Both cuBLAS and MAGMA provide stride-based batched GEMM kernels.
\begin{figure}[!htb]
\centering
\includegraphics[width=0.85\linewidth]{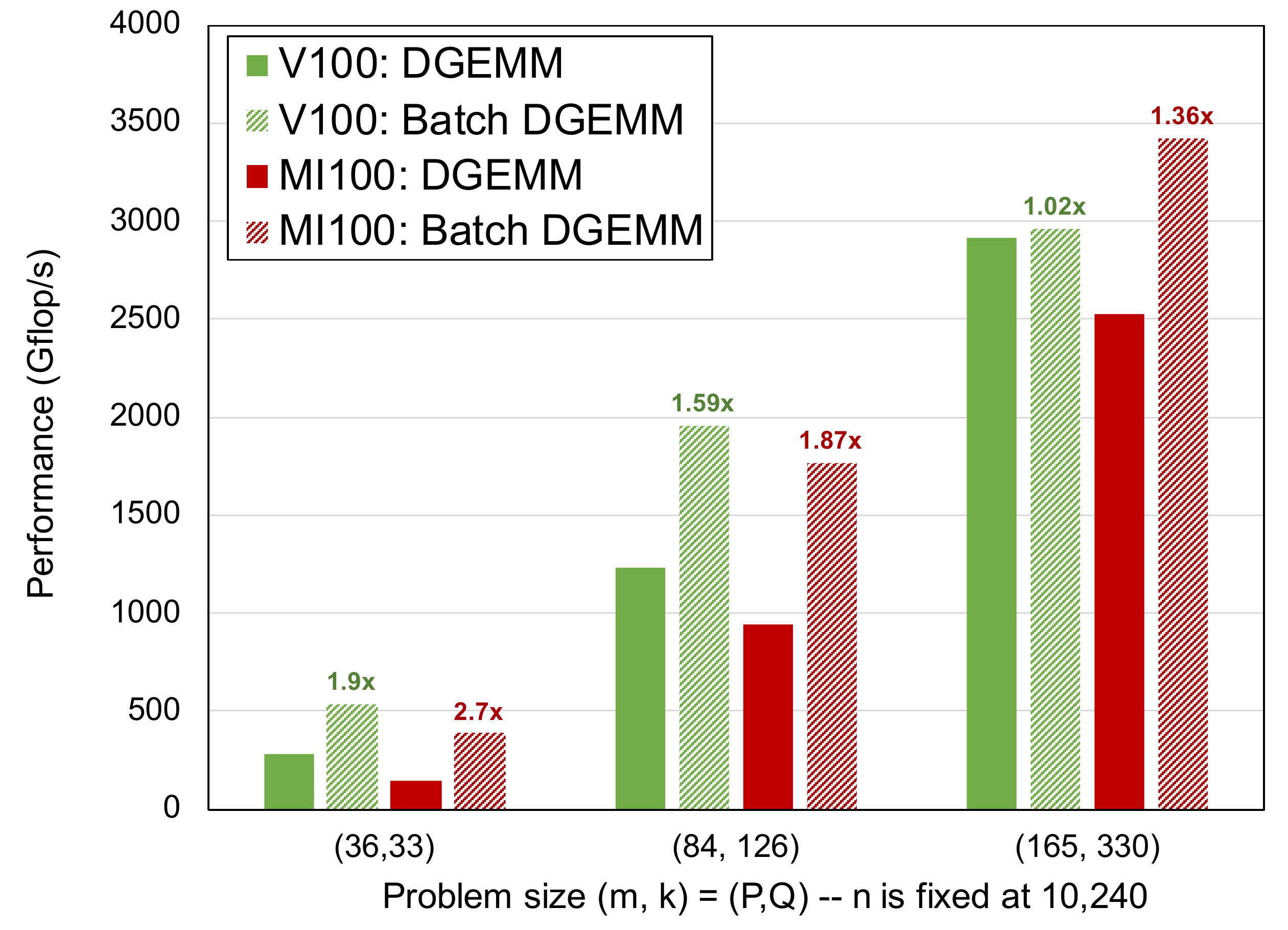}
\vspace{-2mm}
\caption{\reva{Performance of different DGEMM configurations using hipBLAS
  and cuBLAS. Results are for different ($P$, $Q$) pairs on a Nvidia
  V100 (CUDA 11.2) and an AMD Instinct\textsuperscript{TM} MI100 (ROCm 4.2)
  GPUs.}}
\label{fig:magma:dgemm_perf}
\end{figure}

Figure~\ref{fig:magma:dgemm_perf} shows example performance numbers for the GEMM
versus the batched GEMM for typical sizes encountered \reva{in the MFEM-libCEED BP3 \cite{ceed_bp_paper_2020}
 benchmark for triangle or tetrahedron non-tensor elements.
The figure considers an NVIDIA V100 GPU and initial experience with an AMD MI100
GPU. The best-performing routine of MAGMA and the vendor-provided BLAS is shown
for each GPU.
We see that batching the GEMM operation across the $n$
dimension achieves a better performance than launching a single GEMM operation
for these sizes, with the speedup relative to the single GEMM indicated above
each batched GEMM bar.}

To determine the best possible combination of routine (vendor BLAS or MAGMA
library) and batch size $\eta$ (with a standard, non-batch call corresponding to
$\eta = n$), we use data from offline benchmark parameter sweeps to construct a
lightweight abstraction layer. This layer automatically selects the best choice
for the non-tensor GEMM operations. In \cite{magma_libceed2020}, we show that
for higher orders of basis functions, the benefit of the optimized GEMM
formulation is clear in comparison to the pure CUDA kernels in the ``cuda-ref''
backend (up to $10\times$ speedup for the MAGMA formulation on the V100 GPU).

\subsection{MFEM}
MFEM \cite{mfem} is a general-purpose finite element library that since version
4.0 supports hardware accelerators, such as GPUs, as well as programming models
and libraries, such as CUDA, HIP, OCCA \cite{occa},
libCEED \cite{libceed-joss-paper}, RAJA \cite{RAJA-report} and OpenMP.  The goal
of the MFEM developments in CEED is to provide state-of-the-art optimized
performance high-order kernels to applications in an ease of use, flexible
form.

The MFEM performance portability approach is based on a system of backends and
kernels working seamlessly with a lightweight memory spaces manager. A
distinctive feature of this approach is the ability to select the backends at
runtime. For instance, different MPI ranks can choose different backends (like
CPU or GPU), allowing applications to take full advantage of heterogeneous
architectures. Another important aspect of MFEM's approach is the ability to
easily mix CPU-only code with code that utilizes the new backends, thus allowing
for selective gradual transition of existing capabilities. Most of the kernels
are based on a single source, while still offering good performance. For
performance-critical kernels, where a single source does not provide good
performance, the implementation introduces dispatch points based on the selected
backend and, in some cases, on kernel parameters such as the finite element
order.

\begin{figure}[!htb]
\centering
\includegraphics[width=1.0\linewidth]{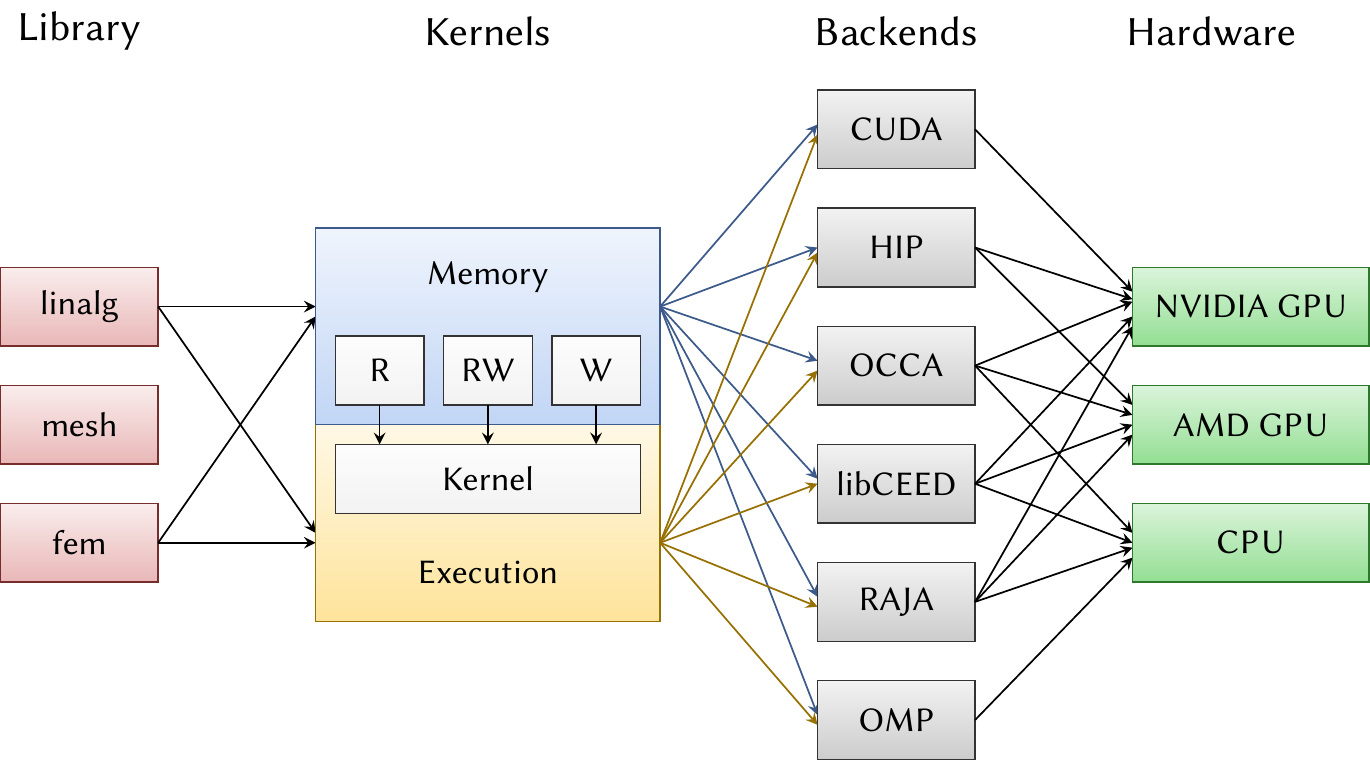}
\vspace{-6mm}
\caption{\label{fig:mfem-gpu-design} Diagram of MFEM's modular design for
  accelerator support, combining flexible memory management with
  runtime-selectable backends for executing key finite element and linear
  algebra kernels.}
\end{figure}

Figure \ref{fig:mfem-gpu-design} illustrates the main components of MFEM's
modular design for accelerator support. The \emph{Library} side of MFEM (on the
left) represents the software components where new kernels have been added.
\emph{Kernels} and \emph{memory management} are the two ingredients most
programming models have to deal with when providing such an abstraction to
address code portability for HPC platforms.
\reva{Similarly to Figure \ref{fig:libceed:libCEEDBackends} the MFEM
design allows for a variety of runtime-selectable backends that can execute its
kernels on both CPU and GPU hardware. Unlike the libCEED figure though,
Figure \ref{fig:mfem-gpu-design} includes the kernel abstraction for a much
broader range of meshing, finite element and linear algebra features that a
general finite element library like MFEM needs to support.}  For more details,
see \cite{mfem,ceed_bp_paper_2020} and Section \ref{sec:discussion}.

\begin{figure}[!htb]
\centering
\includegraphics[width=1.0\linewidth]{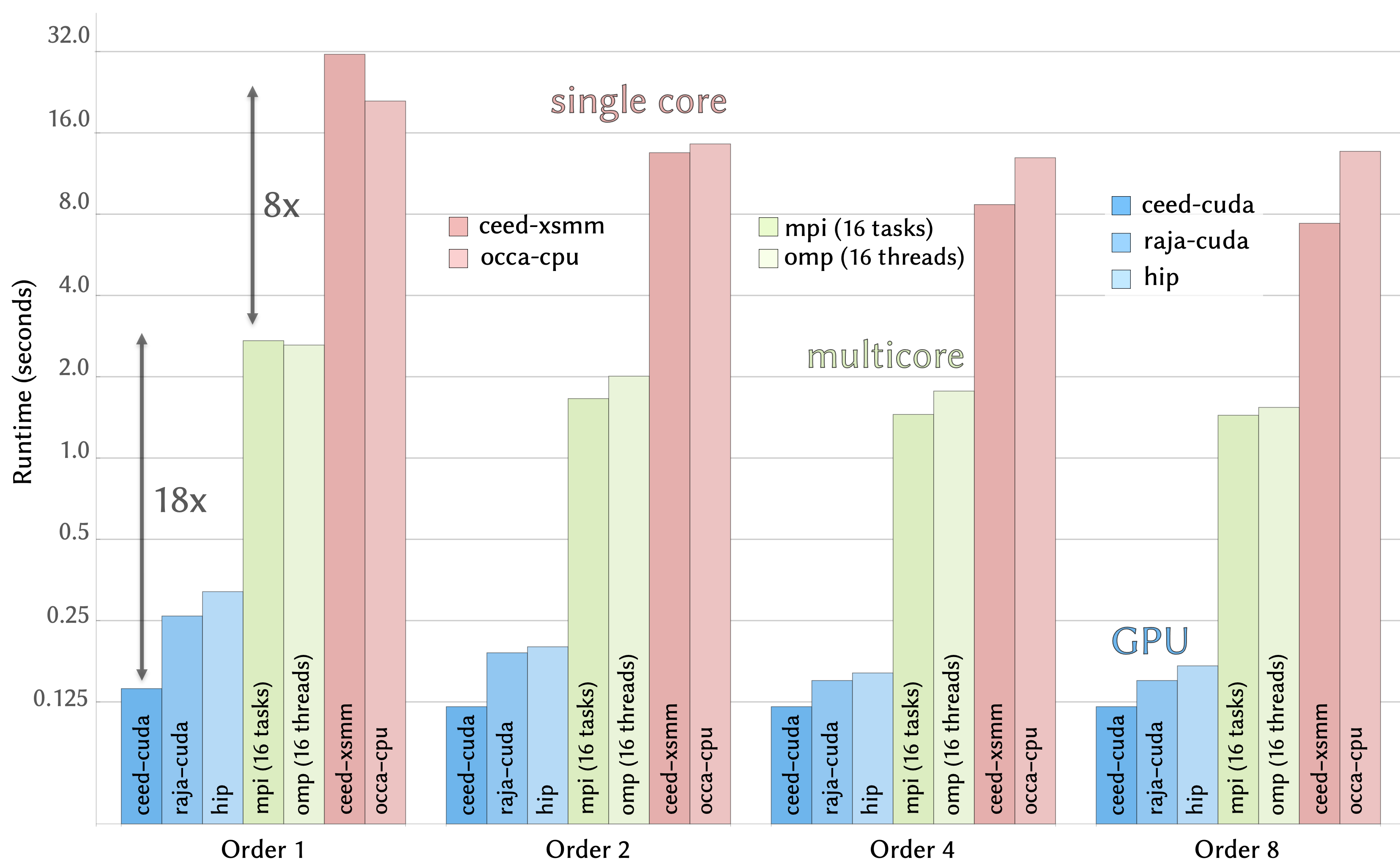}
\vspace{-6mm}
\caption{\label{fig:mfem-gpu-results} Performance results with selection of the
  backend available in MFEM v4.2: 2D Poisson problem with 1.3 million degrees of
  freedom \reva{solved using 200 unpreconditioned CG iterations}, using Intel
  Xeon Gold 6130@2.1GHz CPU plus NVIDIA GV100 (ceed-raja, raja-cuda, CUDA 10.1)
  and AMD Radeon Instinct\textsuperscript{TM} MI60 GPUs (hip, ROCm 3.8).
  Switching from serial to parallel execution on a desktop  workstation leads to
  an order of magnitude performance improvement (note: y-axis is logarithmic).
  Using the desktop GPU results in another order of magnitude performance.
  \reva{Note that these results are representative for the state of the MFEM
  backends as of version 4.2. We do not claim that they represent a fair
  comparison between CPUs and GPUs because not all backends are fully optimized.
  (For example much better CPU results are reported in
  \cite{KronbichlerWall2018} and \cite{KronbichlerLjungkvist2019}.)}
}
\end{figure}

MFEM's GPU acceleration has demonstrated excellent performance in a number of
single-GPU and multi-GPU benchmarks. The high-order algorithms in MFEM are
particularly well suited for GPUs as shown by the results in Figure
\ref{fig:mfem-gpu-results} which report performance results from LLNL's Corona
machine and a Linux configuration similar to the compute nodes of LLNL's Sierra
supercomputer. We note that hand-tuning is still required for good performance
and the AMD results are preliminary.

\subsection{libParanumal}
The Paranumal project \citep{ChalmersKarakusAustinSwirydowiczWarburton2020}
started at Virginia Tech in 2017 as a new GPU effort targeting 90\% of the
capabilities of the CPU version of Nek5000. It was not originally designed as a
user facing library but rather as a collection of high-order finite element
streaming benchmarks (streamParanumal), CEED benchmarks (benchParanumal), and
self contained mini-apps (libParanumal). These were all developed ab initio
using the Open Concurrent Computing Abstraction (OCCA) \citep{occa} and kernel
language (OKL) \citep{medina2015okl}. Although OKL is a generic portable kernel
programming language the initial kernels were optimized for NVIDIA P100 and V100
GPUs, see \cite{swirydowicz2019acceleration}. \reva{By design the libParanumal
OKL kernels use intrinsic C types without recourse to structs and classes
facilitating their use in other projects.}

Many finite element operations are heavily memory bound including Krylov
updates, finite element gather and scatter operations that admit high throughput
hardware agnostic implementations for both the NVIDIA and AMD
GPUs \citep{chalmers2020portable}. Implementations of these kernels have been
released in the streamParanumal standalone benchmark
suite \citep{streamParanumal2020}.

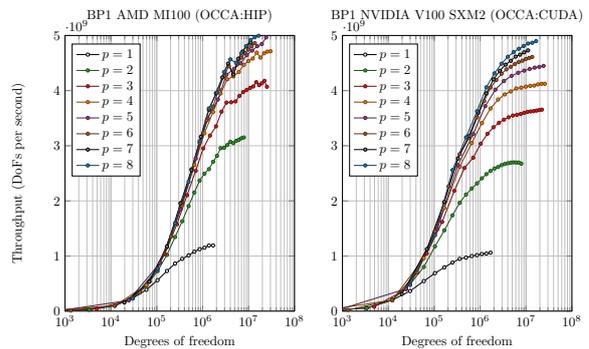
\begin{figure}[!htb]
\centering
\begin{tikzpicture}[scale=0.5]

\begin{axis}[
  xmode=log,
  grid=both,
  major grid style={line width=.1pt,draw=gray!50},
  minor grid style={line width=.1pt,draw=gray!50},
  domain=1:8,
  width=3in,
    height=3.5in,
  ymin=1e-16,
  xlabel={Degrees of freedom},
  ylabel={Throughput (DoFs per second)},
  cycle list name=will,
  legend cell align=left,
  legend pos=north west,
  mark size=1.2pt,
  legend entries={$p=1$,$p=2$,$p=3$,$p=4$,$p=5$,$p=6$,$p=7$,$p=8$},
  title={BP1 AMD MI100 (OCCA:HIP)},
  ymax=5e9,
  ymin=1e6,
  xmin=1e3,
  xmax=1e8,
]
\addplot table[x index=0, y index=1] {
343 2.962494e+06 2.020490e+00 
3375 2.854793e+07 1.692613e+01 
12167 1.000257e+08 5.683243e+01 
29791 2.325910e+08 1.294609e+02 
59319 3.881041e+08 2.134161e+02 
103823 5.621445e+08 3.066587e+02 
166375 7.286035e+08 3.952193e+02 
250047 8.632943e+08 4.663057e+02 
357911 9.485338e+08 5.106738e+02 
493039 1.011586e+09 5.432012e+02 
658503 1.078537e+09 5.779212e+02 
857375 1.124495e+09 6.014818e+02 
1092727 1.150796e+09 6.146303e+02 
1367631 1.191745e+09 6.356870e+02 
1685159 1.191708e+09 6.349643e+02 
};

\addplot table[x index=0, y index=1] {
343 2.993313e+06 9.046631e-01 
3375 2.832288e+07 7.802939e+00 
12167 1.017955e+08 2.730115e+01 
29791 2.434772e+08 6.447200e+01 
59319 4.582831e+08 1.204518e+02 
103823 7.256423e+08 1.897944e+02 
166375 1.020902e+09 2.661026e+02 
250047 1.345443e+09 3.497989e+02 
357911 1.631087e+09 4.232246e+02 
493039 1.905985e+09 4.937751e+02 
658503 2.152251e+09 5.568592e+02 
857375 2.367953e+09 6.120160e+02 
1092727 2.494030e+09 6.440224e+02 
1367631 2.576265e+09 6.647464e+02 
1685159 2.709297e+09 6.986078e+02 
2048383 2.817490e+09 7.260842e+02 
2460375 2.956327e+09 7.614737e+02 
2924207 2.963308e+09 7.629254e+02 
3442951 3.014110e+09 7.756901e+02 
4019679 3.071329e+09 7.901274e+02 
4657463 3.048509e+09 7.839982e+02 
5359375 3.087694e+09 7.938380e+02 
6128487 3.108627e+09 7.990014e+02 
6967871 3.136055e+09 8.058496e+02 
7880599 3.150152e+09 8.092857e+02 
};

\addplot table[x index=0, y index=1] {
125 1.098837e+06 2.682287e-01 
4913 4.127514e+07 8.812140e+00 
24389 2.020759e+08 4.227264e+01 
68921 5.389180e+08 1.118181e+02 
148877 1.035496e+09 2.139051e+02 
274625 1.576488e+09 3.247615e+02 
456533 2.102531e+09 4.323109e+02 
704969 2.546400e+09 5.228581e+02 
1030301 2.952681e+09 6.056479e+02 
1442897 3.184884e+09 6.527413e+02 
1953125 3.352719e+09 6.866846e+02 
2571353 3.618727e+09 7.407630e+02 
3307949 3.783034e+09 7.740438e+02 
4173281 3.789632e+09 7.750930e+02 
5177717 3.804466e+09 7.778672e+02 
6331625 3.903862e+09 7.979581e+02 
7645373 3.973795e+09 8.120458e+02 
9129329 4.013107e+09 8.198943e+02 
10793861 4.048911e+09 8.270431e+02 
12649337 4.075646e+09 8.323542e+02 
14706125 4.154221e+09 8.482637e+02 
16974593 4.099277e+09 8.369213e+02 
19465109 4.133176e+09 8.437293e+02 
22188041 4.177958e+09 8.527665e+02 
25153757 4.069999e+09 8.306375e+02 
};

\addplot table[x index=0, y index=1] {
343 2.902720e+06 5.886558e-01 
12167 1.006193e+08 1.895232e+01 
59319 4.721099e+08 8.788930e+01 
166375 1.160570e+09 2.150302e+02 
357911 1.923687e+09 3.555042e+02 
658503 2.651500e+09 4.892165e+02 
1092727 3.226399e+09 5.946304e+02 
1685159 3.612793e+09 6.653072e+02 
2460375 3.948393e+09 7.266635e+02 
3442951 4.206290e+09 7.737539e+02 
4657463 4.258029e+09 7.829669e+02 
6128487 4.338235e+09 7.974597e+02 
7880599 4.467594e+09 8.210180e+02 
9938375 4.537002e+09 8.335829e+02 
12326391 4.582093e+09 8.417022e+02 
15069223 4.689650e+09 8.613125e+02 
18191447 4.599616e+09 8.446499e+02 
21717639 4.677202e+09 8.587833e+02 
25672375 4.707845e+09 8.643076e+02 
30080231 4.714525e+09 8.654420e+02 
};

\addplot table[x index=0, y index=1] {
729 6.071098e+06 1.119302e+00 
24389 1.950257e+08 3.429213e+01 
117649 8.764754e+08 1.529255e+02 
328509 1.845201e+09 3.209300e+02 
704969 2.768564e+09 4.807013e+02 
1295029 3.498957e+09 6.068629e+02 
2146689 3.839667e+09 6.654632e+02 
3307949 4.291180e+09 7.433147e+02 
4826809 4.394451e+09 7.608906e+02 
6751269 4.495905e+09 7.782056e+02 
9129329 4.644271e+09 8.036769e+02 
12008989 4.726796e+09 8.177816e+02 
15438249 4.805906e+09 8.313185e+02 
19465109 4.845723e+09 8.380775e+02 
24137569 4.965206e+09 8.586290e+02 
};

\addplot table[x index=0, y index=1] {
1331 1.106477e+07 1.926991e+00 
12167 9.695100e+07 1.645185e+01 
42875 3.371027e+08 5.677848e+01 
103823 7.616624e+08 1.278355e+02 
205379 1.355716e+09 2.270728e+02 
357911 1.960748e+09 3.279700e+02 
571787 2.574499e+09 4.302227e+02 
857375 3.050327e+09 5.093790e+02 
1225043 3.461235e+09 5.776828e+02 
1685159 3.711519e+09 6.191875e+02 
2248091 4.026689e+09 6.715300e+02 
2924207 4.268690e+09 7.116802e+02 
3723875 4.443533e+09 7.406474e+02 
4657463 4.314645e+09 7.190127e+02 
5735339 4.487102e+09 7.476152e+02 
6967871 4.599118e+09 7.661566e+02 
8365427 4.685696e+09 7.804699e+02 
9938375 4.745197e+09 7.902823e+02 
11697083 4.806111e+09 8.003381e+02 
13651919 4.862928e+09 8.097186e+02 
};

\addplot table[x index=0, y index=1] {
19683 1.569963e+08 2.582623e+01 
68921 5.340199e+08 8.735004e+01 
166375 1.173135e+09 1.913733e+02 
328509 1.957920e+09 3.188921e+02 
571787 2.558295e+09 4.162468e+02 
912673 3.163633e+09 5.143628e+02 
1367631 3.654786e+09 5.938952e+02 
1953125 3.954985e+09 6.424076e+02 
2685619 4.240232e+09 6.885104e+02 
3581577 4.491418e+09 7.290988e+02 
4657463 4.292093e+09 6.965847e+02 
5929741 4.578204e+09 7.428775e+02 
7414875 4.709931e+09 7.641277e+02 
9129329 4.803064e+09 7.791277e+02 
11089567 4.871738e+09 7.901704e+02 
};

\addplot table[x index=0, y index=1] {
29791 2.288005e+08 3.680740e+01 
103823 7.705576e+08 1.234023e+02 
250047 1.595311e+09 2.549338e+02 
493039 2.347096e+09 3.745978e+02 
857375 3.084656e+09 4.919055e+02 
1367631 3.677302e+09 5.860710e+02 
2048383 3.963071e+09 6.313406e+02 
2924207 4.338772e+09 6.909593e+02 
4019679 4.567294e+09 7.271572e+02 
5359375 4.503633e+09 7.168651e+02 
6967871 4.680298e+09 7.448505e+02 
8869743 4.808083e+09 7.650699e+02 
11089567 4.919258e+09 7.826577e+02 
13651919 4.967295e+09 7.902109e+02 
16581375 4.995556e+09 7.946280e+02 
};

\end{axis}

\end{tikzpicture}
\begin{tikzpicture}[scale=0.5]

\begin{axis}[
  xmode=log,
  grid=both,
  major grid style={line width=.1pt,draw=gray!50},
  minor grid style={line width=.1pt,draw=gray!50},
  domain=1:8,
  width=3in,
    height=3.5in,
  ymin=1e-16,
  xlabel={Degrees of freedom},
  cycle list name=will,
  legend cell align=left,
  legend pos=north west,
  mark size=1.2pt,
  legend entries={$p=1$,$p=2$,$p=3$,$p=4$,$p=5$,$p=6$,$p=7$,$p=8$},
  title={BP1 NVIDIA V100 SXM2 (OCCA:CUDA)},
  ymax=5e9,
  ymin=1e6,
  xmin=1e3,
  xmax=1e8,
]
\addplot table[x index=0, y index=1] {
343 5.668872e+06 3.866303e+00 
3375 5.632795e+07 3.339696e+01 
12167 1.999377e+08 1.136003e+02 
29791 3.654971e+08 2.034369e+02 
59319 5.449371e+08 2.996576e+02 
103823 6.895619e+08 3.761669e+02 
166375 7.939148e+08 4.306463e+02 
250047 8.820929e+08 4.764597e+02 
357911 9.479289e+08 5.103482e+02 
493039 9.709167e+08 5.213625e+02 
658503 9.993494e+08 5.354892e+02 
857375 1.016692e+09 5.438189e+02 
1092727 1.036154e+09 5.534009e+02 
1367631 1.045662e+09 5.577652e+02 
1685159 1.061047e+09 5.653456e+02 
};

\addplot table[x index=0, y index=1] {
343 5.434596e+06 1.642487e+00 
3375 5.449979e+07 1.501466e+01 
12167 2.044883e+08 5.484297e+01 
29791 4.867079e+08 1.288787e+02 
59319 8.016301e+08 2.106946e+02 
103823 1.174711e+09 3.072498e+02 
166375 1.466332e+09 3.822059e+02 
250047 1.738509e+09 4.519912e+02 
357911 1.963851e+09 5.095681e+02 
493039 2.115382e+09 5.480228e+02 
658503 2.226204e+09 5.759933e+02 
857375 2.317726e+09 5.990346e+02 
1092727 2.409727e+09 6.222533e+02 
1367631 2.484768e+09 6.411376e+02 
1685159 2.544191e+09 6.560343e+02 
2048383 2.586653e+09 6.665960e+02 
2460375 2.615149e+09 6.735951e+02 
2924207 2.648711e+09 6.819300e+02 
3442951 2.668793e+09 6.868216e+02 
4019679 2.683807e+09 6.904337e+02 
4657463 2.694186e+09 6.928754e+02 
5359375 2.699128e+09 6.939385e+02 
6128487 2.695963e+09 6.929355e+02 
6967871 2.694305e+09 6.923362e+02 
7880599 2.675707e+09 6.873992e+02 
};

\addplot table[x index=0, y index=1] {
4913 7.994187e+07 1.706739e+01 
24389 3.990594e+08 8.347997e+01 
68921 1.046655e+09 2.171665e+02 
148877 1.618841e+09 3.344080e+02 
274625 2.183746e+09 4.498588e+02 
456533 2.597377e+09 5.340584e+02 
704969 2.783288e+09 5.714989e+02 
1030301 3.037141e+09 6.229722e+02 
1442897 3.216418e+09 6.592043e+02 
1953125 3.335926e+09 6.832451e+02 
2571353 3.417030e+09 6.994752e+02 
3307949 3.473360e+09 7.106815e+02 
4173281 3.517051e+09 7.193419e+02 
5177717 3.547942e+09 7.254179e+02 
6331625 3.561382e+09 7.279544e+02 
7645373 3.576701e+09 7.308994e+02 
9129329 3.598099e+09 7.351064e+02 
10793861 3.614590e+09 7.383273e+02 
12649337 3.622979e+09 7.399078e+02 
14706125 3.628690e+09 7.409538e+02 
16974593 3.638303e+09 7.428074e+02 
19465109 3.647906e+09 7.446683e+02 
22188041 3.654947e+09 7.460143e+02 
};

\addplot table[x index=0, y index=1] {
343 5.777232e+06 1.171591e+00 
12167 1.995780e+08 3.759186e+01 
59319 9.609962e+08 1.789017e+02 
166375 1.873513e+09 3.471243e+02 
357911 2.617179e+09 4.836639e+02 
658503 3.050779e+09 5.628857e+02 
1092727 3.411537e+09 6.287515e+02 
1685159 3.676480e+09 6.770352e+02 
2460375 3.838523e+09 7.064428e+02 
3442951 3.931294e+09 7.231679e+02 
4657463 3.985396e+09 7.328350e+02 
6128487 4.039955e+09 7.426294e+02 
7880599 4.053705e+09 7.449569e+02 
9938375 4.072039e+09 7.481553e+02 
12326391 4.086234e+09 7.506159e+02 
15069223 4.092548e+09 7.516472e+02 
18191447 4.113838e+09 7.554440e+02 
21717639 4.123982e+09 7.572063e+02 
25672375 4.124238e+09 7.571638e+02 
};

\addplot table[x index=0, y index=1] {
729 1.153873e+07 2.127345e+00 
24389 4.042797e+08 7.108607e+01 
117649 1.491839e+09 2.602929e+02 
328509 2.646791e+09 4.603482e+02 
704969 3.212401e+09 5.577639e+02 
1295029 3.718820e+09 6.449962e+02 
2146689 4.017718e+09 6.963216e+02 
3307949 4.175055e+09 7.231997e+02 
4826809 4.267365e+09 7.388858e+02 
6751269 4.323887e+09 7.484305e+02 
9129329 4.362360e+09 7.548930e+02 
12008989 4.393027e+09 7.600364e+02 
15438249 4.417152e+09 7.640725e+02 
19465109 4.430704e+09 7.662991e+02 
24137569 4.448967e+09 7.693563e+02 
};

\addplot table[x index=0, y index=1] {
1331 2.199353e+07 3.830297e+00 
12167 2.022114e+08 3.431375e+01 
42875 6.983720e+08 1.176274e+02 
103823 1.449908e+09 2.433488e+02 
205379 2.263802e+09 3.791708e+02 
357911 2.846663e+09 4.761551e+02 
571787 3.139850e+09 5.246980e+02 
857375 3.516272e+09 5.871879e+02 
1225043 3.809688e+09 6.358400e+02 
1685159 4.055862e+09 6.766336e+02 
2248091 4.191770e+09 6.990606e+02 
2924207 4.305598e+09 7.178336e+02 
3723875 4.385727e+09 7.310122e+02 
4657463 4.433121e+09 7.387561e+02 
5735339 4.483129e+09 7.469534e+02 
6967871 4.520119e+09 7.529963e+02 
8365427 4.543637e+09 7.568079e+02 
9938375 4.573896e+09 7.617532e+02 
11697083 4.593222e+09 7.648867e+02 
13651919 4.611214e+09 7.678060e+02 
};

\addplot table[x index=0, y index=1] {
19683 3.108887e+08 5.114187e+01 
68921 1.125552e+09 1.841075e+02 
166375 1.981620e+09 3.232613e+02 
328509 2.767968e+09 4.508270e+02 
571787 3.176236e+09 5.167886e+02 
912673 3.601426e+09 5.855418e+02 
1367631 3.950021e+09 6.418703e+02 
1953125 4.203461e+09 6.827676e+02 
2685619 4.359858e+09 7.079348e+02 
3581577 4.471059e+09 7.257939e+02 
4657463 4.541343e+09 7.370368e+02 
5929741 4.617515e+09 7.492564e+02 
7414875 4.652674e+09 7.548385e+02 
9129329 4.695769e+09 7.617229e+02 
11089567 4.729538e+09 7.671062e+02 
};

\addplot table[x index=0, y index=1] {
29791 4.744191e+08 7.632036e+01 
103823 1.378185e+09 2.207118e+02 
250047 2.560169e+09 4.091201e+02 
493039 3.153188e+09 5.032507e+02 
857375 3.625750e+09 5.781931e+02 
1367631 4.060580e+09 6.471561e+02 
2048383 4.313066e+09 6.870969e+02 
2924207 4.488815e+09 7.148541e+02 
4019679 4.614497e+09 7.346723e+02 
5359375 4.698353e+09 7.478597e+02 
6967871 4.758205e+09 7.572491e+02 
8869743 4.809612e+09 7.653132e+02 
11089567 4.854060e+09 7.722847e+02 
13651919 4.867993e+09 7.744136e+02 
16581375 4.897527e+09 7.790348e+02 
};

\end{axis}

\end{tikzpicture}
\vspace{-2mm}
\caption {Performance of the benchParanumal version of the CEED benchmark problem
BP1 on a single GPU of HPE/Tulip AMS Instinct\textsuperscript{TM} MI100 (left) and NVIDIA V100 SXM2 (right).}
\label{libp-bp1-mi100.fig}
\end{figure}

\begin{figure}[!htb]
\centering
\begin{tikzpicture}[scale=0.5]

\begin{axis}[
  xmode=log,
  grid=both,
  major grid style={line width=.1pt,draw=gray!50},
  minor grid style={line width=.1pt,draw=gray!50},
  domain=1:8,
  width=3in,
  height=3.5in,
  ymin=1e-16,
  xlabel={Degrees of freedom},
  ylabel={Throughput (DoFs per second)},
  cycle list name=will,
  legend cell align=left,
  legend pos=north west,
  mark size=1.2pt,
  legend entries={$p=1$,$p=2$,$p=3$,$p=4$,$p=5$,$p=6$,$p=7$,$p=8$},
  title={BP5 AMD MI100 (OCCA:HIP)},
  ymax=4e9,
  ymin=0,
  xmin=1e3,
  xmax=1e7
]

\addplot table[x index=0, y index=1] {
343 2.929675e+06 3.022825e+00 
3375 2.885989e+07 2.531820e+01 
12167 9.972270e+07 8.320963e+01 
29791 2.190766e+08 1.784026e+02 
59319 3.663593e+08 2.940783e+02 
103823 5.271976e+08 4.191816e+02 
166375 6.816318e+08 5.383315e+02 
250047 7.925976e+08 6.228256e+02 
357911 8.375667e+08 6.555997e+02 
493039 9.028205e+08 7.044807e+02 
658503 9.555374e+08 7.437245e+02 
857375 1.004632e+09 7.802860e+02 
1092727 1.041560e+09 8.075245e+02 
1367631 1.075999e+09 8.329515e+02 
1685159 1.095090e+09 8.466089e+02 
};

\addplot table[x index=0, y index=1] {
343 2.925921e+06 1.435699e+00 
3375 2.911154e+07 1.248070e+01 
12167 1.024391e+08 4.216803e+01 
29791 2.417859e+08 9.759999e+01 
59319 4.507689e+08 1.798773e+02 
103823 6.846292e+08 2.711370e+02 
166375 9.382678e+08 3.695994e+02 
250047 1.195060e+09 4.688768e+02 
357911 1.442340e+09 5.641481e+02 
493039 1.608240e+09 6.274884e+02 
658503 1.674403e+09 6.519910e+02 
857375 1.691570e+09 6.575755e+02 
1092727 1.874324e+09 7.275914e+02 
1367631 1.969268e+09 7.635248e+02 
1685159 2.039717e+09 7.900137e+02 
2048383 2.104073e+09 8.141961e+02 
2460375 2.151562e+09 8.319030e+02 
2924207 2.177010e+09 8.411412e+02 
3442951 2.213839e+09 8.548249e+02 
4019679 2.235025e+09 8.625100e+02 
4657463 2.205183e+09 8.505522e+02 
5359375 2.226508e+09 8.583723e+02 
6128487 2.222494e+09 8.564559e+02 
6967871 2.253437e+09 8.680380e+02 
7880599 2.276245e+09 8.765059e+02 
};

\addplot table[x index=0, y index=1] {
125 1.105090e+06 4.543382e-01 
4913 4.223619e+07 1.386357e+01 
24389 2.012332e+08 6.363009e+01 
68921 5.275488e+08 1.642753e+02 
148877 9.584580e+08 2.959800e+02 
274625 1.423685e+09 4.373477e+02 
456533 1.856615e+09 5.682919e+02 
704969 2.149533e+09 6.562265e+02 
1030301 2.285231e+09 6.962606e+02 
1442897 2.444264e+09 7.435439e+02 
1953125 2.603007e+09 7.908284e+02 
2571353 2.724574e+09 8.268962e+02 
3307949 2.797620e+09 8.483202e+02 
4173281 2.860974e+09 8.668835e+02 
5177717 2.850618e+09 8.631906e+02 
6331625 2.898732e+09 8.772691e+02 
7645373 2.920533e+09 8.834333e+02 
9129329 2.951080e+09 8.922856e+02 
10793861 2.989156e+09 9.034482e+02 
12649337 3.018807e+09 9.120932e+02 
14706125 3.044976e+09 9.197118e+02 
16974593 3.055090e+09 9.225046e+02 
19465109 3.060262e+09 9.238275e+02 
22188041 3.096226e+09 9.344634e+02 
25153757 3.100681e+09 9.356049e+02 
};

\addplot table[x index=0, y index=1] {
343 2.910178e+06 9.515880e-01 
12167 1.038356e+08 2.937366e+01 
59319 4.583760e+08 1.264783e+02 
166375 1.071306e+09 2.925734e+02 
357911 1.754318e+09 4.764158e+02 
658503 2.255086e+09 6.102406e+02 
1092727 2.521749e+09 6.807391e+02 
1685159 2.761396e+09 7.441051e+02 
2460375 3.058898e+09 8.231545e+02 
3442951 3.132083e+09 8.419488e+02 
4657463 3.163272e+09 8.495991e+02 
6128487 3.245031e+09 8.709383e+02 
7880599 3.301427e+09 8.855458e+02 
9938375 3.344073e+09 8.965290e+02 
12326391 3.380333e+09 9.058539e+02 
15069223 3.431850e+09 9.193095e+02 
18191447 3.475565e+09 9.307084e+02 
21717639 3.492132e+09 9.348684e+02 
25672375 3.500002e+09 9.367283e+02 
30080231 3.498739e+09 9.361687e+02 
};

\addplot table[x index=0, y index=1] {
729 6.272355e+06 1.806565e+00 
24389 2.028694e+08 5.264212e+01 
117649 8.242394e+08 2.099861e+02 
328509 1.681227e+09 4.250551e+02 
704969 2.355796e+09 5.931150e+02 
1295029 2.610274e+09 6.554507e+02 
2146689 2.985169e+09 7.482261e+02 
3307949 3.175818e+09 7.949548e+02 
4826809 3.246437e+09 8.118089e+02 
6751269 3.351768e+09 8.374795e+02 
9129329 3.443306e+09 8.597967e+02 
12008989 3.494014e+09 8.719944e+02 
15438249 3.550064e+09 8.855875e+02 
19465109 3.611612e+09 9.005989e+02 
24137569 3.597380e+09 8.967565e+02 
};

\addplot table[x index=0, y index=1] {
1331 1.146328e+07 3.048028e+00 
12167 1.032599e+08 2.581276e+01 
42875 3.481652e+08 8.541358e+01 
103823 7.271448e+08 1.767741e+02 
205379 1.212792e+09 2.932649e+02 
357911 1.719967e+09 4.144406e+02 
571787 2.155516e+09 5.180930e+02 
857375 2.474738e+09 5.937125e+02 
1225043 2.553449e+09 6.117122e+02 
1685159 2.784491e+09 6.662942e+02 
2248091 3.037750e+09 7.262137e+02 
2924207 3.139200e+09 7.498809e+02 
3723875 3.210740e+09 7.664645e+02 
4657463 3.219874e+09 7.682113e+02 
5735339 3.311877e+09 7.897761e+02 
6967871 3.333851e+09 7.946768e+02 
8365427 3.364715e+09 8.017320e+02 
9938375 3.404945e+09 8.110471e+02 
11697083 3.427870e+09 8.162639e+02 
13651919 3.461619e+09 8.240789e+02 
};

\addplot table[x index=0, y index=1] {
19683 1.475910e+08 3.534951e+01 
68921 5.036424e+08 1.187930e+02 
166375 1.109371e+09 2.597261e+02 
328509 1.672513e+09 3.898535e+02 
571787 2.338659e+09 5.435509e+02 
912673 2.773712e+09 6.433417e+02 
1367631 2.872856e+09 6.653157e+02 
1953125 3.131738e+09 7.244077e+02 
2685619 3.476433e+09 8.033775e+02 
3581577 3.532272e+09 8.156500e+02 
4657463 3.565788e+09 8.228596e+02 
5929741 3.663263e+09 8.448938e+02 
7414875 3.732541e+09 8.604715e+02 
9129329 3.766381e+09 8.679232e+02 
11089567 3.840886e+09 8.847804e+02 
};

\addplot table[x index=0, y index=1] {
29791 1.702837e+08 3.952151e+01 
103823 7.189810e+08 1.647320e+02 
250047 1.508028e+09 3.433554e+02 
493039 2.094917e+09 4.752135e+02 
857375 2.648586e+09 5.993363e+02 
1367631 2.900236e+09 6.551388e+02 
2048383 3.224750e+09 7.274972e+02 
2924207 3.501558e+09 7.891486e+02 
4019679 3.615638e+09 8.142037e+02 
5359375 3.658434e+09 8.233002e+02 
6967871 3.736657e+09 8.404445e+02 
8869743 3.811581e+09 8.569006e+02 
11089567 3.865956e+09 8.687816e+02 
13651919 3.899537e+09 8.760286e+02 
16581375 3.935818e+09 8.839148e+02 
};

\end{axis}

\end{tikzpicture}
\begin{tikzpicture}[scale=0.5]

\begin{axis}[
  xmode=log,
  grid=both,
  major grid style={line width=.1pt,draw=gray!50},
  minor grid style={line width=.1pt,draw=gray!50},
  domain=1:8,
  width=3in,
  height=3.5in,
  ymin=1e-16,
  xlabel={Degrees of freedom},
  cycle list name=will,
  legend cell align=left,
  legend pos=north west,
  mark size=1.2pt,
  legend entries={$p=1$,$p=2$,$p=3$,$p=4$,$p=5$,$p=6$,$p=7$,$p=8$},
  title={BP5 NVIDIA V100 SXM2 (OCCA:CUDA)},
  ymax=4e9,
  ymin=0,
  xmin=1e3,
  xmax=1e7
]

\addplot table[x index=0, y index=1] {
343 5.464528e+06 5.638275e+00 
3375 5.339989e+07 4.684665e+01 
12167 1.910241e+08 1.593925e+02 
29791 3.494295e+08 2.845540e+02 
59319 5.075519e+08 4.074143e+02 
103823 6.471278e+08 5.145396e+02 
166375 7.389421e+08 5.835934e+02 
250047 8.208868e+08 6.450553e+02 
357911 8.809631e+08 6.895679e+02 
493039 9.116402e+08 7.113628e+02 
658503 9.318642e+08 7.252989e+02 
857375 9.484539e+08 7.366529e+02 
1092727 9.636873e+08 7.471494e+02 
1367631 9.757287e+08 7.553301e+02 
1685159 9.803936e+08 7.579380e+02 
};

\addplot table[x index=0, y index=1] {
343 5.364480e+06 2.632258e+00 
3375 5.553245e+07 2.380788e+01 
12167 1.961114e+08 8.072725e+01 
29791 4.621196e+08 1.865405e+02 
59319 6.927135e+08 2.764242e+02 
103823 1.008956e+09 3.995815e+02 
166375 1.198254e+09 4.720124e+02 
250047 1.413365e+09 5.545275e+02 
357911 1.558040e+09 6.094024e+02 
493039 1.649969e+09 6.437697e+02 
658503 1.718375e+09 6.691132e+02 
857375 1.776506e+09 6.905931e+02 
1092727 1.824193e+09 7.081308e+02 
1367631 1.865487e+09 7.232869e+02 
1685159 1.888495e+09 7.314429e+02 
2048383 1.907519e+09 7.381370e+02 
2460375 1.919154e+09 7.420422e+02 
2924207 1.932901e+09 7.468238e+02 
3442951 1.946006e+09 7.514070e+02 
4019679 1.956685e+09 7.550968e+02 
4657463 1.961195e+09 7.564444e+02 
5359375 1.969089e+09 7.591312e+02 
6128487 1.969994e+09 7.591532e+02 
6967871 1.972327e+09 7.597526e+02 
7880599 1.967076e+09 7.574553e+02 
};

\addplot table[x index=0, y index=1] {
125 1.931221e+06 7.939872e-01 
4913 7.109652e+07 2.333666e+01 
24389 3.878626e+08 1.226424e+02 
68921 9.013895e+08 2.806870e+02 
148877 1.308814e+09 4.041730e+02 
274625 1.793310e+09 5.508943e+02 
456533 2.028946e+09 6.210409e+02 
704969 2.165607e+09 6.611336e+02 
1030301 2.300818e+09 7.010093e+02 
1442897 2.383135e+09 7.249486e+02 
1953125 2.439765e+09 7.412335e+02 
2571353 2.477729e+09 7.519799e+02 
3307949 2.508020e+09 7.605052e+02 
4173281 2.529836e+09 7.665477e+02 
5177717 2.551977e+09 7.727598e+02 
6331625 2.566366e+09 7.766824e+02 
7645373 2.576834e+09 7.794674e+02 
9129329 2.589334e+09 7.829085e+02 
10793861 2.593497e+09 7.838636e+02 
12649337 2.600063e+09 7.855751e+02 
14706125 2.604474e+09 7.866614e+02 
16974593 2.609757e+09 7.880335e+02 
19465109 2.615055e+09 7.894290e+02 
22188041 2.616889e+09 7.897959e+02 
25153757 2.619663e+09 7.904617e+02 
};

\addplot table[x index=0, y index=1] {
343 5.217969e+06 1.706204e+00 
12167 1.852950e+08 5.241739e+01 
59319 7.985170e+08 2.203324e+02 
166375 1.542433e+09 4.212380e+02 
357911 2.053131e+09 5.575635e+02 
658503 2.346951e+09 6.350997e+02 
1092727 2.535631e+09 6.844864e+02 
1685159 2.664582e+09 7.180168e+02 
2460375 2.740410e+09 7.374487e+02 
3442951 2.797176e+09 7.519208e+02 
4657463 2.829747e+09 7.600202e+02 
6128487 2.857222e+09 7.668536e+02 
7880599 2.868452e+09 7.694082e+02 
9938375 2.877695e+09 7.714956e+02 
12326391 2.885708e+09 7.733055e+02 
15069223 2.892820e+09 7.749163e+02 
18191447 2.900426e+09 7.766942e+02 
21717639 2.904671e+09 7.776008e+02 
25672375 2.908222e+09 7.783462e+02 
30080231 2.912893e+09 7.794121e+02 
};

\addplot table[x index=0, y index=1] {
729 1.117602e+07 3.218919e+00 
24389 3.858582e+08 1.001255e+02 
117649 1.267481e+09 3.229078e+02 
328509 2.123454e+09 5.368607e+02 
704969 2.459434e+09 6.192076e+02 
1295029 2.728848e+09 6.852252e+02 
2146689 2.863661e+09 7.177705e+02 
3307949 2.950311e+09 7.385070e+02 
4826809 2.999701e+09 7.501099e+02 
6751269 3.026364e+09 7.561735e+02 
9129329 3.052696e+09 7.622609e+02 
12008989 3.067907e+09 7.656517e+02 
15438249 3.075462e+09 7.671947e+02 
19465109 3.084405e+09 7.691334e+02 
24137569 3.092555e+09 7.709136e+02 
};

\addplot table[x index=0, y index=1] {
1331 2.089382e+07 5.555558e+00 
12167 1.926539e+08 4.815933e+01 
42875 6.759285e+08 1.658221e+02 
103823 1.160714e+09 2.821779e+02 
205379 1.808039e+09 4.372014e+02 
357911 2.201736e+09 5.305268e+02 
571787 2.433757e+09 5.849702e+02 
857375 2.617833e+09 6.280424e+02 
1225043 2.785840e+09 6.673845e+02 
1685159 2.909293e+09 6.961577e+02 
2248091 2.967483e+09 7.094153e+02 
2924207 3.022778e+09 7.220704e+02 
3723875 3.068723e+09 7.325623e+02 
4657463 3.096508e+09 7.387781e+02 
5735339 3.121408e+09 7.443553e+02 
6967871 3.141814e+09 7.489016e+02 
8365427 3.154819e+09 7.517189e+02 
9938375 3.168079e+09 7.546262e+02 
11697083 3.178880e+09 7.569730e+02 
13651919 3.185026e+09 7.582327e+02 
};

\addplot table[x index=0, y index=1] {
19683 3.093509e+08 7.409263e+01 
68921 9.519400e+08 2.245319e+02 
166375 1.705179e+09 3.992168e+02 
328509 2.249358e+09 5.243128e+02 
571787 2.635148e+09 6.124608e+02 
912673 2.905921e+09 6.740065e+02 
1367631 3.060563e+09 7.087862e+02 
1953125 3.192269e+09 7.384092e+02 
2685619 3.267165e+09 7.550173e+02 
3581577 3.321730e+09 7.670329e+02 
4657463 3.362020e+09 7.758370e+02 
5929741 3.387298e+09 7.812454e+02 
7414875 3.409557e+09 7.860133e+02 
9129329 3.419976e+09 7.880977e+02 
11089567 3.443886e+09 7.933280e+02 
};

\addplot table[x index=0, y index=1] {
29791 4.477943e+08 1.039295e+02 
103823 1.218459e+09 2.791719e+02 
250047 1.892990e+09 4.310054e+02 
493039 2.349629e+09 5.329927e+02 
857375 2.728839e+09 6.174963e+02 
1367631 2.982990e+09 6.738323e+02 
2048383 3.118776e+09 7.035895e+02 
2924207 3.195612e+09 7.201972e+02 
4019679 3.278775e+09 7.383457e+02 
5359375 3.326937e+09 7.486995e+02 
6967871 3.353244e+09 7.542077e+02 
8869743 3.384054e+09 7.607862e+02 
11089567 3.398976e+09 7.638390e+02 
13651919 3.415253e+09 7.672344e+02 
16581375 3.427125e+09 7.696713e+02 
};

\end{axis}

\end{tikzpicture}
\vspace{-2mm}
\caption{Performance of the benchParanumal version of the CEED benchmark problem
BP5 on a single GPU of HPE/Tulip AMD Instinct\textsuperscript{TM} MI100 (left) and NVIDIA V100 SXM2 (right).}
\label{libp-bp5-mi100.fig}
\end{figure}

Portable implementations of the CEED BP benchmarks \cite{ceed_bp_paper_2020} are
available in the benchParanumal library. Figures \ref{libp-bp1-mi100.fig}
and \ref{libp-bp5-mi100.fig} show the performance of the benchParanumal
implementations for the CEED BP1 and BP5 benchmark respectively on a single AMD
MI100 (ROCm 3.9) and NVIDIA V100 SXM2 (CUDA 10.1). The benchmarks achieved sustained average memory
bandwidth depending on polynomial degree and mesh size of up to 950GB/s (AMD
MI100) and 800GB/s (NVIDIA V100 SXM2) \reva{despite involving sequences of tensor contractions for each element in the matrix-vector operations.}

The libParanumal library is a self contained high-order finite element library
that uses the same highly optimized OKL kernels as the streamParanumal and
benchParnaumal benchmark suites. It also includes sub-libraries for dense linear
algebra, Krylov solvers, parallel mesh handling and polynomial approximation,
p-type and algebraic multigrid, time stepping, gather-scatter operations and
halo exchanges, and core miscellaneous operations. The libParanumal
sub-libraries support meshes consisting of triangles, quadrilaterals,
tetrahedra, or hexes.
The libParanumal project also includes mini-apps demonstrating GPU accelerated
PDE solvers for linearized acoustics, scalar advection, Galerkin-Boltzmann
finite moment gas-dynamics, compressible Navier-Stokes, elliptic equations,
Fokker-Planck, and incompressible Navier-Stokes. Each solver supports multi-GPU
simulation via Nek's gslib for efficient MPI based gather-scatter operations and
halo exchanges \citep{gsLib}.

The libParanumal library is highly modular and
early fork of the project has been used with modifications as a platform for the custom physics
requirements of the Nek5000 user community \cite{nekrs2020} as described in the next section
albeit without the most recent developments in portable streaming and operator
kernels. The algorithms and design principles of libParanumal kernels have also
influenced the design of kernels produced by the libCEED \texttt{cuda-gen}
backend (see Section \ref{subsec:libceed}). \reva{libParanumal is also beginning to be used
in non-CEED projects as for example in high fidelity time-dependent room acoustic modeling \cite{melander2020massive}.}

\subsection{Nek5000/RS}
\reva{
Nek5000 \cite{nek} is a spectral element code that is used for a wide range
of thermal-fluids applications.  A companion code, NekCEM \cite{nekcem},
is used for computational electromagnetics. These codes have scaled
to millions of MPI ranks using the Nek-based {\em gsLib} communication library
to handle all near-neighbor and other stencil type communications (e.g., for
algebraic multigrid)~\cite{fischer2015scaling}. On CPUs, tensor contractions
constitute the principal computational kernel (typically $>$ 90\% of the
flops). These can be cast as small dense matrix-matrix products resulting in
high performance with a minor amount of tuning \cite{dfm02}.

For GPU-based platforms, node-level parallelism requires kernels written at a
higher level than simple tensor contractions. Early GPU ports started with
NekCEM~\cite{min_2016_a}, using OpenACC and running on OLCF/Titan up to 16,384
NVIDIA K20X GPUs. The OpenACC-based implementation and performance studies were
extended to Nek5000~\cite{gong16,otero19}.

For portability and performance reasons, we decided to develop a new version
of Nek5000, called NekRS, which is written in C++/OCCA \cite{occa}. The NekRS
kernels started as a fork from libParanumal \reva{in late 2018} and were
tailored and expanded to meet the specific requirements of large-scale
turbulent flow simulations in complex domains (e.g., as illustrated in
Figure~\ref{fig:neksolve}). It retains access to the standard Nek5000
interface, which allows users to leverage existing user-specific source code
such as statistical analysis tools for turbulence.
}

Several recent developments in NekRS have led to significant performance gains
on Summit. These include {\em (i)} An accelerator-oriented variant of {\em
gslib} \cite{gsLib} that selects from several communication strategies, including
pack/unpack on the host or device, and GPU-direct or host-based communication.
Runtime-adaptation picks the fastest strategy.
{\em (ii)} Chebyshev-accelerated additive Schwarz smoothing.
This approach combines the standard Nek5000 additive Schwarz method with the
Chebyshev-accelerated Jacobi smoothing provided in libParanumal \reva{(and, e.g.,
deal.ii \cite{KronbichlerLjungkvist2019})}.  Local Schwarz
solves are performed with fast-diagonalization implemented with tensor
contractions that have a complexity that is on par with operator evaluation.
{\em (iii)} Projection-based initial guesses to avoid redundant iteration work
in successive timesteps \cite{fisc98,austin2020initial}. The performance impact
of these developments are described in detail in \cite{nekrs2020}.

\begin{table} [t]
  \footnotesize
  \begin{center} \begin{tabular}{|c|c|c|c|c|}
  \hline
   System & Device & Backend  & $t_{step} (s) $ &  R \\
  \hline
       Summit        &  V100   & CUDA & {\tt 8.51e-02} & {\tt 1}    \\
       Tulip         &  MI100  & HIP  & {\tt 9.96e-02} & {\tt 0.85} \\
       Tulip         &  MI60   & HIP  & {\tt 1.41e-01} & {\tt 0.60} \\
       Tulip         &  V100   & CUDA & {\tt 8.85e-02} & {\tt 0.96} \\
       Theta-GPU     &  A100   & CUDA & {\tt 5.59e-02} & {\tt 1.52} \\
  \hline
  \end{tabular}
\end{center}
\vspace{-2mm}
\caption{NekRS Navier-Stokes performance on a single GPU of HPE/Tulip AMD
  Instinct\textsuperscript{TM} MI100, AMD Radeon
  Instinct\textsuperscript{TM} MI60, NVIDIA V100 PCIe and ALCF/Theta-GPU
  NVIDIA A100 SXM4 , compared to OLCF/Summit NVIDIA V100 SXM2,
  for turbulent pipe flow simulation with $Re= 19,000$, $E= 6840$, $p=6$,
  and $n=2,346,120$.
  Time per step in seconds ($t_{step}$) is averaged over 100 steps.
  R is the ratio of $t_{step}$ on Summit V100 to that on other systems.}
\label{pipe-baseline}
\end{table}

\reva{
The advantage of basing NekRS on OCCA is clear from the results of
Table~\ref{pipe-baseline}, which demonstrates full Navier-Stokes performance
results \revb{for} NVIDIA and AMD GPUs.  The table shows a single-GPU
comparison of the averaged-walltime per timestep for the MI60, MI100, and A100,
compared to a single V100 on Summit. We remark that extensive tuning has been
applied for the V100, which has been the primary development platform for
NekRS.  Despite this, the performance on the other GPUs is within the scope of
what we would expect for these nodes.  The A100 performs remarkably well, with
speedup 1.5$\times$ of the performance of the V100 and at a
near-strong-scale-limit value of $n=2.22$M gridpoints on a single GPU.  Despite
the fact that the AMD GPU code has seen less tuning than the NVIDIA code the
fact that the performance is on par illustrates the portability provided by the
OCCA base.
}

\section{Discussion} \label{sec:discussion}

In this section we share some of the porting experiences on the CEED project,
and discuss some of the GPU lessons we have learned.

Overall, we have found that porting to GPU architectures is a disruptive
process, similar to the transition from serial to MPI parallel programming. As
such, we recommend to start a new code for the GPU port, if possible, (as with
NekRS and libParanumal) as opposed to incrementally porting an existing code
(as with MFEM). \reva{We also recommend taking advantage of GPU-accelerated libraries,
such as libCEED, when applicable.}
For low-order applications, the CEED software also provides access to
new classes of algorithms (high-order methods) that can take better advantage of
GPU hardware compared to traditional low-order approaches
\cite{ceed_bp_paper_2020}.

For new codes, we have found the use of OCCA and its OKL language a pragmatic
choice that has allowed us to make quick progress in capturing e.g. the
capabilities of Nek5000 in libParanumal without choosing a specific
manufacturer's GPU programming model since OCCA translates OKL code into CUDA,
HIP, OpenCL, or OpenMP at runtime for native just-in-time (JIT)
compilation. OCCA has also been instrumental in the exploration of the
high-order algorithmic space, as different versions of the CEED kernels can be
easily implemented, modified and tested with it. \reva{Examples are provided with the OCCA distribution that demonstrate the simplicity of the API and kernel language \cite{occa-web}.}

For the porting of existing codes, we have found that integration of kernels at
the \emph{for-loop} level, as with Kokkos and RAJA, has several important
benefits. For example, in MFEM, the original code was transformed to use a new
\emph{for-loop} abstraction defined as a set of \code{MFEM\_FORALL} macros, in
order to take advantage of various \emph{backends} supported via the new
macros. This approach allows for gradual code transformations that are not too
disruptive for both MFEM developers and users. Existing applications based on
MFEM are able to continue to work as before with easy transition to accelerated
kernels. This approach also allows interoperability with other software
components and external libraries that can be used in conjunction with MFEM
(e.g., \emph{hypre}, PETSc, SUNDIALS). The main challenge in this transition to
{\em kernel-centric} implementation is the need to transform existing algorithms
to take full advantage of the increased levels of parallelism in the
accelerators while maintaining good performances on standard CPU architectures.

An important aspect of GPU programming is the need to manage memory allocation
and transfers between the CPU (host) and the accelerator (device). This can be a
frequent source of bugs and inefficiencies in complex applications. For example
in MFEM, a special \code{Memory} class was introduced to manage a pair of host
and device pointers and provides a simple interface for copying or moving the
data when needed. An important feature of this \revb{\texttt{\small Memory}} class is the ability to work with
externally allocated host and/or device pointers which is essential for
interoperability with other libraries. The \code{Memory} has also been useful in
the porting of MFEM-based applications, see Section \ref{sec_marbl}.

Finally, the optimization of GPU kernels really requires understanding of the
GPU hardware and its multi-level memory hierarchy as well as advanced techniques
such as code generation and JIT compilation. To illustrate these points, we
describe in the rest of the section the sequence of developments that led to the
\texttt{cuda-gen} backend of libCEED, which achieves close to peak performance
for high-order operator evaluation with tensor finite elements on NVIDIA GPUs.

The first libCEED CUDA backend was the reference backend \texttt{cuda-ref},
which established a blueprint for GPU porting in libCEED. The main difficulty at
this point was to handle all the runtime aspects of libCEED. Being able to
produce efficient GPU kernels relies on the compiler knowing as much as possible
during compilation, which conflicts with the generality of the libCEED approach
where users are free to specify e.g., polynomial order and number of quadrature
points used at runtime. For this reason, it was critical to use JIT compilation
to generate on the fly GPU kernels with as much information as we could provide
at runtime. In general, we believe that JIT compilation will play an important
role in HPC in the future, and is essentially a requirement for high-order
applications with runtime order selection.

\begin{figure*}[!htb]
\centering
\begin{tikzpicture}[scale=0.75]

\begin{axis}[
  xmode=log,
  grid=both,
  major grid style={line width=.1pt,draw=gray!50},
  minor grid style={line width=.1pt,draw=gray!50},
  domain=1:8,
  width=3in,
  height=3in,
  ymin=1e-16,
  xlabel={Degrees of freedom},
  ylabel={Throughput (DoFs per second)},
  cycle list name=will,
  legend cell align=left,
  legend pos=north west,
  mark size=1.2pt,
  legend entries={$p=1$,$p=2$,$p=3$,$p=4$,$p=5$,$p=6$,$p=7$,$p=8$},
  title={cuda-ref},
  ymax=3e9,
  ymin=0,
  xmin=1e3,
  xmax=5e7,
]
\addplot
table {%
8 0
12 0
18 0
27 97697.6
45 194075
75 408615
125 681725
225 1317300
405 2397200
729 4326020
1377 7345500
2601 13406900
4913 21571900
9537 37171000
18513 54262200
35937 71510700
70785 84305200
139425 94766400
274625 98994700
545025 104644000
1081665 107214000
2146689 107340000
};
\addplot
table {%
27 86642.3
45 190571
75 354171
125 627263
225 1294040
405 2463230
729 4053390
1377 8164550
2601 13358500
4913 23295200
9537 35613100
18513 63801900
35937 106972000
70785 173311000
139425 242215000
274625 302885000
545025 333711000
1081665 355903000
2146689 361575000
4276737 360498000
};
\addplot
table {%
64 206320
112 538437
196 963521
343 1817400
637 3703260
1183 6181430
2197 12506000
4225 21799400
8125 36595300
15625 64729900
30625 88808000
60025 158292000
117649 222210000
232897 321802000
461041 434068000
912673 523288000
1815937 585137000
3613153 574914000
7189057 582601000
};
\addplot
table {%
125 568968
225 1174340
405 2145150
729 3843590
1377 7938780
2601 12469300
4913 22466200
9537 43376000
18513 66784200
35937 92031500
70785 116525000
139425 190024000
274625 307060000
545025 445833000
1081665 576265000
2146689 694312000
4276737 714708000
8520321 630608000
16974593 621652000
};
\addplot
table {%
216 827571
396 1723660
726 3104560
1331 6137470
2541 11389600
4851 20985100
9261 37704800
18081 62762600
35301 93443500
68921 124183000
136161 136117000
269001 233118000
531441 352774000
1056321 494411000
2099601 586572000
4173281 704987000
8320641 709309000
16589601 621744000
};
\addplot
table {%
343 1307130
637 2684010
1183 5006050
2197 7850920
4225 15665900
8125 31868700
15625 50965800
30625 92106100
60025 120712000
117649 142243000
232897 148352000
461041 259554000
912673 378420000
1815937 511265000
3613153 618792000
7189057 676405000
14340865 657659000
};
\addplot
table {%
512 1543500
960 3179690
1800 5989380
3375 10988500
6525 20858600
12615 38787300
24389 69297300
47937 110819000
94221 142767000
185193 171396000
367137 183620000
727833 290478000
1442897 382343000
2873025 476871000
5720625 547777000
11390625 534553000
22730625 601257000
};
\addplot
table {%
729 1712120
1377 3821430
2601 7183580
4913 12801300
9537 24551100
18513 46399400
35937 70931700
70785 108381000
139425 138587000
274625 147895000
545025 164207000
1081665 235497000
2146689 338028000
4276737 421840000
8520321 476137000
16974593 456454000
};
\end{axis}

\end{tikzpicture}
\begin{tikzpicture}[scale=0.75]

\begin{axis}[
  xmode=log,
  grid=both,
  major grid style={line width=.1pt,draw=gray!50},
  minor grid style={line width=.1pt,draw=gray!50},
  domain=1:8,
  width=3in,
  height=3in,
  ymin=1e-16,
  xlabel={Degrees of freedom},
  cycle list name=will,
  legend cell align=left,
  legend pos=north west,
  mark size=1.2pt,
  legend entries={$p=1$,$p=2$,$p=3$,$p=4$,$p=5$,$p=6$,$p=7$,$p=8$},
  title={cuda-shared},
  ymax=3e9,
  ymin=0,
  xmin=1e3
]
\addplot
table {%
8 0
12 0
18 0
27 76265.4
45 174252
75 299402
125 572952
225 1138880
405 2179690
729 3927580
1377 7684420
2601 13003900
4913 23902100
9537 41634300
18513 64217300
35937 83605200
70785 105491000
139425 122397000
274625 120511000
545025 129618000
1081665 133112000
2146689 133070000
};
\addplot
table {%
27 72436.6
45 160631
75 299335
125 584540
225 1207870
405 2024440
729 4028960
1377 7017900
2601 14684400
4913 26660800
9537 50464000
18513 97273200
35937 150242000
70785 242245000
139425 318346000
274625 377641000
545025 422649000
1081665 442673000
2146689 456147000
4276737 466545000
};
\addplot
table {%
64 171035
112 444494
196 878765
343 1634680
637 3574880
1183 6640390
2197 11062800
4225 24029500
8125 45920300
15625 83056200
30625 145701000
60025 247565000
117649 395873000
232897 521449000
461041 601109000
912673 654095000
1815937 709137000
3613153 720960000
7189057 734091000
};
\addplot
table {%
125 527780
225 1083300
405 2236020
729 3988100
1377 7524200
2601 13376800
4913 24952600
9537 54545500
18513 95181500
35937 169099000
70785 320881000
139425 497370000
274625 654293000
545025 740582000
1081665 838356000
2146689 870932000
4276737 910358000
8520321 932327000
16974593 938926000
};
\addplot
table {%
216 1014050
396 2124550
726 4003500
1331 6690320
2541 13628400
4851 27379000
9261 53348800
18081 104514000
35301 169845000
68921 300728000
136161 480716000
269001 665309000
531441 808609000
1056321 895884000
2099601 972577000
4173281 1018490000
8320641 1045770000
16589601 1055500000
};
\addplot
table {%
343 1748910
637 3428130
1183 5946930
2197 8814530
4225 22685000
8125 43056700
15625 79286900
30625 163833000
60025 300633000
117649 454352000
232897 669388000
461041 894676000
912673 1019600000
1815937 1136440000
3613153 1195110000
7189057 1230400000
14340865 1241850000
};
\addplot
table {%
512 2677850
960 5002380
1800 8601500
3375 16999000
6525 31665900
12615 68035300
24389 130586000
47937 222694000
94221 378650000
185193 546904000
367137 799319000
727833 876349000
1442897 974729000
2873025 1072400000
5720625 1101870000
11390625 1130070000
22730625 1144770000
};
\addplot
table {%
729 3323270
1377 6487690
2601 13035600
4913 23371700
9537 45293000
18513 93399400
35937 170833000
70785 328470000
139425 499187000
274625 686939000
545025 878333000
1081665 1007340000
2146689 1093960000
4276737 1168660000
8520321 1197690000
16974593 1221050000
};
\end{axis}

\end{tikzpicture}
\begin{tikzpicture}[scale=0.75]

\begin{axis}[
  xmode=log,
  grid=both,
  major grid style={line width=.1pt,draw=gray!50},
  minor grid style={line width=.1pt,draw=gray!50},
  domain=1:8,
  width=3in,
  height=3in,
  ymin=1e-16,
  xlabel={Degrees of freedom},
  cycle list name=will,
  legend cell align=left,
  legend pos=north west,
  mark size=1.2pt,
  legend entries={$p=1$,$p=2$,$p=3$,$p=4$,$p=5$,$p=6$,$p=7$,$p=8$},
  title={cuda-gen},
  ymax=3e9,
  ymin=0,
  xmin=1e3
]
\addplot
table {%
8 0
12 0
18 0
27 123325
45 264812
75 466405
125 818155
225 1875270
405 3523990
729 5821200
1377 11645300
2601 20269600
4913 35076300
9537 71115500
18513 124823000
35937 180679000
70785 289482000
139425 365728000
274625 400287000
545025 475610000
1081665 477441000
2146689 504134000
4276737 510151000
};
\addplot
table {%
27 115749
45 254339
75 463086
125 832409
225 1670560
405 3158070
729 6375570
1377 10791200
2601 22455700
4913 40320400
9537 75215000
18513 155953000
35937 241249000
70785 432183000
139425 689521000
274625 923184000
545025 1097690000
1081665 1257310000
2146689 1326730000
4276737 1398360000
8520321 1420110000
};
\addplot
table {%
64 294484
112 744010
196 1366060
343 2866360
637 5364720
1183 9148380
2197 19046700
4225 34401100
8125 70765500
15625 113698000
30625 231765000
60025 382212000
117649 686966000
232897 1024820000
461041 1346210000
912673 1556310000
1815937 1705060000
3613153 1810140000
7189057 1871520000
};
\addplot
table {%
125 879152
225 1798400
405 3086970
729 6295260
1377 12383300
2601 20220700
4913 36076100
9537 76611000
18513 153422000
35937 264560000
70785 447091000
139425 745282000
274625 1193790000
545025 1545050000
1081665 1850930000
2146689 1997630000
4276737 2173980000
8520321 2261880000
16974593 2299990000
};
\addplot
table {%
216 1388580
396 2824060
726 5245600
1331 10940100
2541 20643400
4851 39869500
9261 68436600
18081 134737000
35301 249069000
68921 495580000
136161 720951000
269001 1196310000
531441 1501450000
1056321 1806100000
2099601 2023560000
4173281 2240860000
8320641 2339920000
16589601 2344130000
};
\addplot
table {%
343 2468010
637 4425750
1183 9831920
2197 16226900
4225 32892200
8125 60084700
15625 114974000
30625 231110000
60025 418114000
117649 784957000
232897 1228440000
461041 1646680000
912673 1903820000
1815937 2229650000
3613153 2533280000
7189057 2636130000
14340865 2682530000
};
\addplot
table {%
512 3732730
960 7049960
1800 13322600
3375 24668800
6525 54279400
12615 104883000
24389 184793000
47937 363765000
94221 604109000
185193 981768000
367137 1385550000
727833 1720690000
1442897 2131400000
2873025 2283350000
5720625 2477530000
11390625 2420680000
22730625 2639210000
};
\addplot
table {%
729 5270580
1377 10214100
2601 19479100
4913 35650700
9537 76196100
18513 153738000
35937 268224000
70785 454863000
139425 772477000
274625 1242340000
545025 1592660000
1081665 1941970000
2146689 2211030000
4276737 2485130000
8520321 2529180000
16974593 2598970000
33883137 2609470000
};
\end{axis}

\end{tikzpicture}
\vspace{-2mm}
\caption{Performance of the \texttt{cuda-ref} (left) \texttt{cuda-shared}
(center) and \texttt{cuda-gen} (right) backends of libCEED for the CEED benchmark
problem BP3 on an NVIDIA V100 GPU.}
\label{fig:libceed:backends}
\end{figure*}
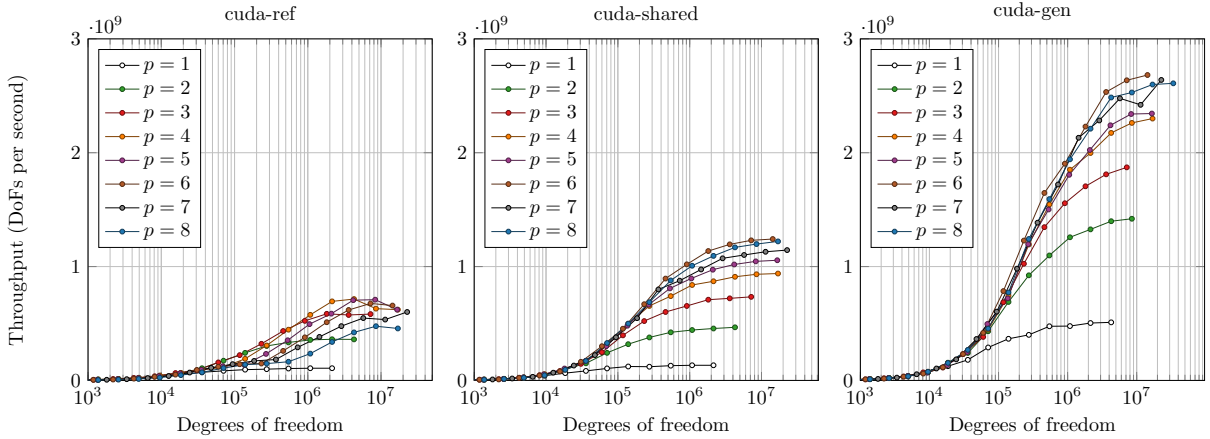

The second libCEED CUDA backend, \texttt{cuda-shared}, focused on
optimizing each GPU kernel individually to achieve peak performance. The tensor
kernels are highly memory bound, so the challenge was to use the different
memory bandwidths efficiently. Typically, the bottlenecks are the local/shared
memory bandwidths and the memory access patterns to the data.
If we compare the \texttt{cuda-ref}
and \texttt{cuda-shared} backends in Figure \ref{fig:libceed:backends}, we see
that for low orders (1 to 3) the performance of the \texttt{cuda-ref} and
\texttt{cuda-shared} backends are similar, the \texttt{cuda-ref} kernels do not
yet saturate the local/shared memory bandwidth. However, for orders higher to 4,
we observe that the \texttt{cuda-ref} backend performance deteriorates with the
order. This is due to local/shared memory bandwidth getting more and more
saturated. On the other hand, the \texttt{cuda-shared} manages by careful memory
accesses and unrolling loops to continue saturating the global memory bandwidth
and thus achieves high performance for each GPU kernel.

The final, and best performing backend, \texttt{cuda-gen} uses a code generation
approach, based on the \texttt{cuda-shared} backend, to generate at runtime
(with JIT compilation) a unique optimized GPU kernel representing the whole
operator. Since the \texttt{cuda-ref}, \texttt{cuda-shared} are decomposing the
matrix free operators in a sequence of GPU kernels, they require the storage, in
global memory, of unnecessary temporary results in between each kernel
launch. Fusing GPU kernels prevents these unnecessary data storage and movements
between kernel launches resulting in a 2-3 time speedup over the
\texttt{cuda-shared} backend, see Figure \ref{fig:libceed:backends}, and around
5 time speedup over the reference backend \texttt{cuda-ref} on the CEED
benchmark problem BP3.

\section{Applications} \label{sec:applications}

\begin{figure*}[t] \centering
\includegraphics[width=1.\textwidth]{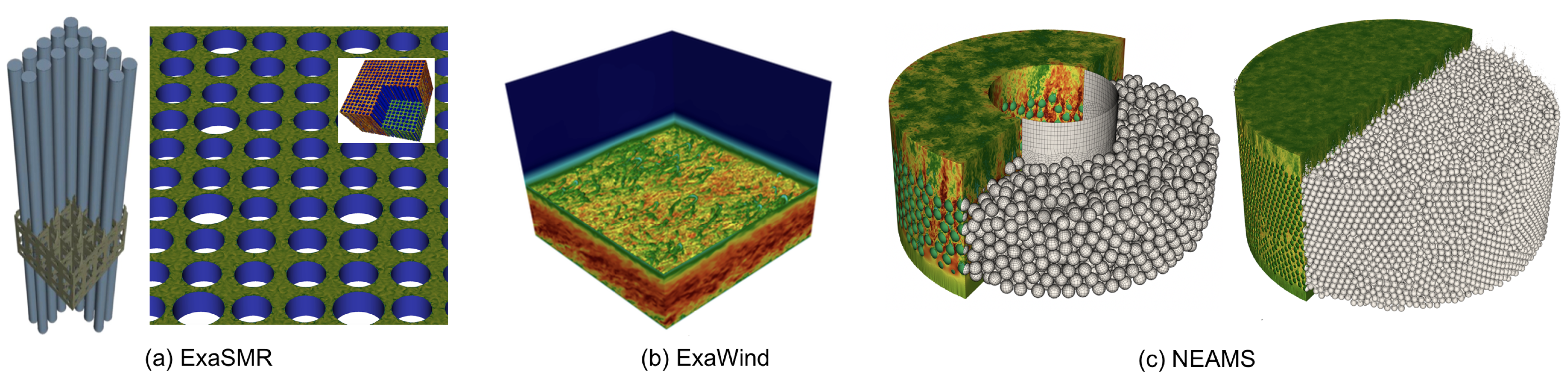}
\vspace{-4mm}
\caption{Nek5000/RS applications: (a) ExaSMR's 17$\times 17$ fuel rod
configuration and turbulent flows profile, (b) ExaWind's atmospheric boundary
layer modeling and analysis, (c) NEAMS's pebble-bed reactor configurations
with simulation demonstrating turbulent flows past 3344 pebbles in an annulus
and 44257 pebbles in a cylinder.}
\label{fig:neksolve}
\end{figure*}

The CEED effort includes the development of algorithms, software, simulation and
modeling, performance analysis and optimization for CEED-engaged applications.
While we are focused on exascale applications in the ECP, CEED is also extending
its contribution to a broader range of engineering and science application
areas, such as nuclear energy, wind energy, fusion, solid mechanics, additive
manufacturing, internal combustion, and recent extension to weather modeling and
aerosol transport research. In this section, we demonstrate the impact of the
CEED-developed open source codes, Nek5000/RS and MFEM with full simulation
capability, scaling on various acceleration architectures (including the full
scaling performance on Summit GPUs), in DOE's ExaSMR, ExaWind, NEAMS, MARBL, and
ExaAM projects.

\subsection{ExaSMR}
\begin{table*} [t]
  \footnotesize
  \begin{center} \begin{tabular}{|c|c|c|c|c|c|c|c|c|c|c|c|c|}
  \hline
  \multicolumn{13}{|c|}{{\bf ExaSMR application performance: $17 \times 17$ fuel rods simulation}} \\
  \hline
  \hline
  case & node & gpu &  $E$ & $N$ & $E$/gpu & $n$/gpu & $v_i$ & $p_i$ &  $t_{step} (s) $ &  R  & R$_{\rm ideal}$ & P$_{\rm eff}$ (\%) \\
  \hline
         &1810 &10860 & 175618000 & 7 &16171 & 5.5M & 4 & 2 & 1.855e-01 &1.00 & 1.00  & 100 \\
  strong &2536 &15216 & 175618000 & 7 &11542 & 3.9M & 4 & 2 & 1.517e-01 &1.22 & 1.40  &  87 \\
         &3620 &21720 & 175618000 & 7 & 8085 & 2.7M & 4 & 2 & 1.120e-01 &1.65 & 2.00  &  82 \\
         &4180 &25080 & 175618000 & 7 & 7002 & 2.4M & 4 & 2 & 1.128e-01 &1.64 & 2.30  &  71 \\
         &4608 &27648 & 175618000 & 7 & 6351 & 2.1M & 4 & 2 & 1.038e-01 &1.78 & 2.54  &  70 \\
  \hline
  \hline
  \hline
  case & node & gpu &  $E$ & $N$ & $E$/gpu & $n$/gpu & $v_i$ & $p_i$ & $t_{step} (s) $ &  R  & R$_{\rm ideal}$ & P$_{\rm eff}$ (\%) \\
  \hline
         &87   & 522   & 3324000  & 7 & 6367 & 2.1M &  4 & 2 & 8.57e-02 & 1.00 & 1.00 & 100 \\
         &320  & 1920  & 12188000 & 7 & 6347 & 2.1M &  4 & 2 & 8.67e-02 & 0.98 & 1.00 & 98 \\
  weak   &800  & 4800  & 30470000 & 7 & 6347 & 2.1M &  4 & 2 & 9.11e-02 & 0.94 & 1.00 & 94 \\
         &1600 & 9600  & 60940000 & 7 & 6347 & 2.1M &  4 & 2 & 9.33e-02 & 0.91 & 1.00 & 91 \\
         &3200 & 19200 & 121880000& 7 & 6347 & 2.1M &  4 & 2 & 9.71e-02 & 0.88 & 1.00 & 88 \\
         &4608 & 27648 & 175618000& 7 & 6351 & 2.1M &  4 & 2 & 1.03e-01 & 0.83 & 1.00 & 83 \\
  \hline
  \end{tabular}
\end{center}
\vspace{-2mm}
\caption{\label{nek-large}
ExaSMR:
NekRS strong and weak scaling performed on Summit, using 6 GPUs per node, for
simulating turbulent flow in the $17 \times 17$ rod-bundle of
Figure~\ref{fig:neksolve}(a), right, with $Re_D= 5000$.
Time per step in seconds ($t_{step}$), velocity
iteration count ($v_i$), and pressure iteration count ($p_i$), are all averaged
over 100 steps.  R is the ratio of $t_{step}$ of 1810 nodes to that of others
for strong scaling and $t_{step}$ of 87 nodes to that of others for weak
scaling, provided with the ideal ratio, R$_{\rm ideal}$ and the parallel
efficiency, P$_{\rm eff}$ .}
\end{table*}
\reva{
ExaSMR's target geometry is a small modular reactor assembly comprising 37
bundles, each having a 17$\times$17 array of rods, which totals to $\sim$10,000
long communicating channels. For development, we consider two geometries: a
very long single 17$\times$17 bundle, and a collection of 37 such bundles that are
shorter in length. We analyzed NekRS performance behaviors for both geometries.
Detailed performance for various geometries is discussed in \cite{nekrs2020}.
Here we present the baseline performance of the long 17$\times$17 bundle,
illustrated in Fig.~\ref{fig:neksolve}(a),
having the ratio between the characteristic length $L$ and the rod diameter
$D$ as $L/D\approx 288$.
}

Table~\ref{nek-large} demonstrates strong- and weak-scaling runs out to 175
million elements on Summit---roughly twelve times larger than 15M-element
that were ``hero calculations'' on Mira as recently as 2020.  We measured the
average-walltime per step in seconds, $t_{step}$, using 101-200 steps for
simulations with $Re_D=5000$.
For the strong scaling, we used $E= 175,618,000$ and $N=7$, totaling 60 billion grid
points.  We observe the 17$\times$17 rod-bundle case continues to scale well
to all of Summit, using $n/P=2.1M$ with 70\% parallel efficiency from the base
of 1810 nodes using $n/P=5.5M$, where $P$ is the number of V100s. We see 80\%
efficiency is sustained for $n/P=2.6M$.  The weak scaling uses the meshes
increased by 120, 440, 1100, 2200, 4400, and 6340 layers in the streamwise ($z$)
direction, extruded from a two-dimensional $17\times 17$ mesh having $E=27,700$
spectral elements.  Weak-scaling this problem from 271 to 4608 nodes (1626 to
27648 GPUs) sustains more than 80\% parallel efficiency throughout, using
$2.1M$ grid points per GPU.

We note that the pressure iteration counts, $p_i$, are relatively very low for the
17$\times$17 bundle compared to the pebble cases, which have $p_i\sim$ 8
for the same timestepper and preconditioner.
The geometric complexity of the 17$\times$17 rod-bundle is relatively mild
compared to the pebble case and also the synthetic initial condition does not
quickly transition to full turbulence.
We expect higher pressure iteration counts (e.g., $p_i \sim$ 4--8) once fully
turbulent flow is established for this case.

\subsection{ExaWind}
Efficient simulation of atmospheric boundary layer flows (ABL) is important
for the study of wind farms, urban canyons, and basic weather modeling. In
collaboration with the ECP ExaWind team, we identified a well-documented test
case, the Global Energy and Water Cycle Experiment Atmospheric Boundary Layer
Study (GABLS), to demonstrate the suitability of high-order methods for large
eddy simulations (LES) of the ABL.  Initial convergence results of an LES study
with Nek5000 are shown in Figure~\ref{fig:neksolve}(b).
We have initiated a performance study for this problem on Theta-GPU (A100s) and
Summit (V100s) for an $E=32768$ spectral element mesh with $N=7$ (i.e.,
$n$=11.2M). Our simulations represent turbulent flows on the physical domain
[400m $\times$ 400m $\times$ 400m] with geostrophic wind speed of 8m/s in
$x$-direction and reference potential temperature of 263.5K with no-slip
boundary as well as wall functions based on log-law at the lower wall,
otherwise periodic boundary conditions.
Single-node scaling shows the 80\% strong-scale limit to be 1.8M points/GPU
for both the V100 and A100, with the A100 running at .055s/step and 1.55 times
faster than the V100.

\subsection{NEAMS}
Figure~\ref{fig:neksolve}(c) demonstrates turbulent flows past 3344 pebbles in
an annulus (left) and 44257 pebbles in a cylinder (right) that were computed
with NekRS on Summit using 840 GPUs and 1788 GPUs, respectively.  The cylinder
case has 13M elements of order $N=7$, for a total of 4.4B grid points.  The
annulus configuration is a prototype for pebble-bed reactor configurations that
are being studied by the DOE's Nuclear Energy Advanced Modeling and Simulation
project.  We have developed a novel meshing strategy for generating
\reva{high-quality hexahedral element meshes that ensure accurate representation of
densely packed spheres for these geometries~\cite{ylan21}}.
The meshing algorithm includes Voronoi
tessellation, edge collapse, facet projection onto the spheres, and mesh
smoothing with quality measurements.
These simulations strong-scale well and the NEAMS target configuration of an
annulus with 300,000 pebbles will require about 30B grid points, which is well
within the current performance envelope on Summit.

\subsection{MARBL} \label{sec_marbl}
MARBL is a next-gen multi-physics simulation code being developed at LLNL. The
code provides multi-material radiation-magneto-hydrodynamics with applications
in inertial confinement fusion (ICF), pulsed power and equation of
state/material strength experiments as part of the NNSA ATDM program.
\reva{
One of the central components of MARBL is the BLAST package
\citep{Anderson2018}, which uses an ALE formulation to simulate
conservation laws of mass, momentum, and energy in a moving material frame.}
The BLAST package utilizes
high-order finite element discretizations of physical processes on a high-order
(curved) moving mesh.
\reva{
BLAST's finite element discretization infrastructure is entirely based on the
MFEM library. Therefore, the GPU port of BLAST makes extensive use of on the
matrix-free approach and GPU support via MFEM.  In this
section we provide specifics about the major GPU kernels in BLAST, and the
impact of the CEED project in these GPU development efforts. }

\vspace{-2mm}
\paragraph{Memory management} Since MARBL/BLAST is based on MFEM, it directly uses
the high-level memory management interface for reading and writing device data.
In addition, the MARBL team has enhanced the MFEM's memory manager capabilities
by introducing the Umpire \citep{umpire} memory manager providing access to
memory pools.  This approach enables the following benefits: substantially
reduces slowdowns caused by \code{cudaMalloc} performance; sharing of device memory
buffers inside MARBL to reduce the total device usage; and sharing overall
temporary memory between other external packages in MARBL that use Umpire.

\vspace{-2mm}
\paragraph{Lagrangian phase} In this phase the multi-material compressible Euler
equations are solved on a moving curved mesh \citep{Dobrev2012, Dobrev2016}.
The optimization of the needed mass and force operators has been aided by the
matrix-free methods that were introduced by the CEED-developed Laghos miniapp
\citep{laghos},
which models the main computational kernels of Lagrangian hydrodynamics.
The GPU kernels for these
methods were implemented by the MARBL team and reside in the BLAST code. The
latest Laghos GPU implementations of these kernels give an alternative that
might be used in the future, based on performance tests. A key CEED benefit
provided to MARBL is the ability to drop in replacements for these expensive
kernels as they become available.
Physics-specific quadrature point computations were implemented by the MARBL
team, making use of the RAJA nested parallel loop abstractions combined with
MFEM's GPU capabilities, including GPU-friendly data structures, small dense
matrix kernels, and use of shared memory.
This phase also requires the computation of a \textit{hyperviscosity}
\cite{hypervisc} coefficient, which involves consecutive applications of a
Laplacian operator.  This procedure has been ported on the GPU by applying
directly the MFEM's optimized diffusion kernels.

\vspace{-2mm}
\paragraph{Remesh phase}
\reva{
The mesh optimization phase of BLAST is based on the Target-Matrix
Optimization Paradigm (TMOP), where the mesh optimization problem is posed as a
variational minimization of a nonlinear functional \cite{TMOP_2019, TMOP_2020}. }
The development of the GPU port was performed in MFEM's mesh optimization
miniapp, and then directly ported to MARBL, as both codes use the same core TMOP
algorithms.

\begin{figure}[h!]
\begin{center}
\includegraphics[width=0.5\textwidth]{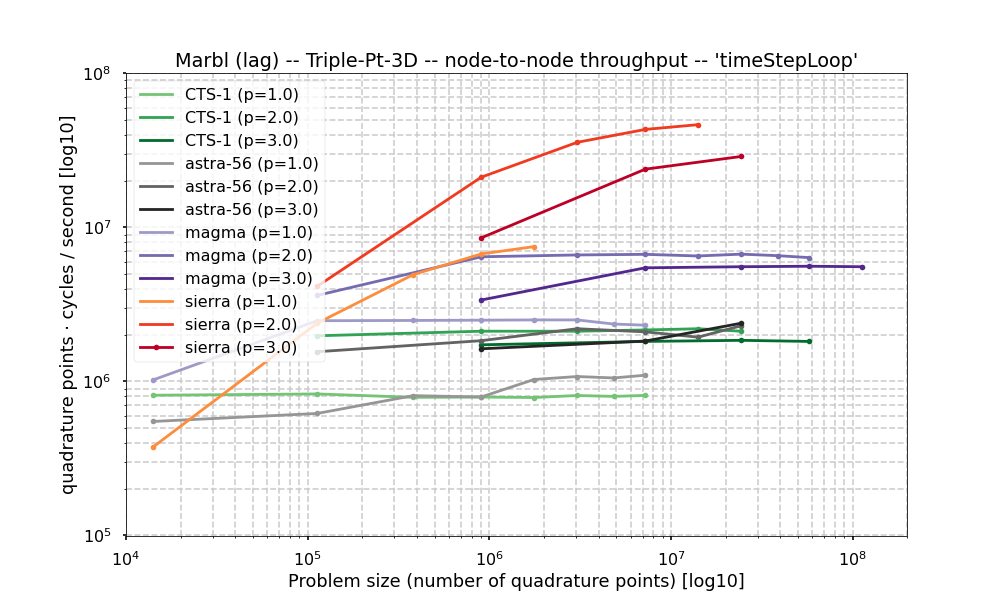}
\vspace{-2mm}
\caption{3D Triple-point problem throughput test in MARBL. Comparison of 3
  CPU-based systems versus the NVIDIA V100 GPU-based Sierra.}
\label{fig_blast}
\end{center}
\end{figure}

\vspace{-2mm}
\paragraph{Remap phase} The remap algorithm in BLAST has two main components,
namely, velocity remap, which is solved by a continuous Galerkin advection
discretization, and remap of other fields, which is modeled by flux-limited
discontinuous Galerkin (DG) advection \citep{Anderson2015, Anderson2016}.
Using the MFEM infrastructure, the MARBL developers have
developed custom GPU code \reva{for matrix-free DG advection remap}.
It is expected that this approach will be improved significantly by the future
work in the CEED-developed Remhos miniapp, as it contains novel matrix-free
DG remap methods \cite{Hajduk2020}.
The continuous Galerkin advection solve is also fully GPU ported.
Similarly to the CG mass matrix inversion in the Lagrangian phase, the remap GPU
code is implemented inside MARBL, and the alternative to switching to the
optimized MFEM kernels will be explored.
In Figure \ref{fig_blast} we present a recent study of MARBL that compares
node-to-node throughput of several CPU machines at LLNL (a Commodity Technology
System (\textit{CTS}), \textit{Astra} and \textit{Magma}) versus the LLNL
\textit{Sierra} machine, showing clear advantage of the GPU executions.  Table
\ref{tab_blast} shows timings of the three main phases of the application, along
with the final speedup of the matrix-free CPU vs GPU kernels.

\begin{table}[h!]
\begin{center}
\resizebox{\columnwidth}{!}{
   \begin{tabular}{c | c c c c}
   \hline
      Phase          & FA CPU    & PA CPU  & PA GPU & speedup  \\
      \hline
      Time Loop & 3854.16 & 2866.54 & 221.03 & 12.9 \\
      Lagrange  & 1773.68 & 1098.42 & 69.73  & 15.7 \\
      Remesh    & 557.98  & 366.24  & 42.67  & 8.5  \\
      Remap     & 1513.99 & 1393.34 & 100.95 & 13.8 \\
   \hline
   \end{tabular} }
\end{center}
\vspace{-2mm}
\caption{CPU and GPU timings on 3 nodes of the LLNL's \textit{rzgenie} (36 tasks
  per node) and \textit{rzansel} (4 GPUs per node) machines. FA is traditional
  full assembly, while PA is matrix-free partial assembly. This is a full 3D
  high-order ALE simulation with 224,160 elements.}
\label{tab_blast}
\end{table}

\subsection{ExaConstit}
\begin{figure}
\centering
\includegraphics[width=0.45\textwidth]{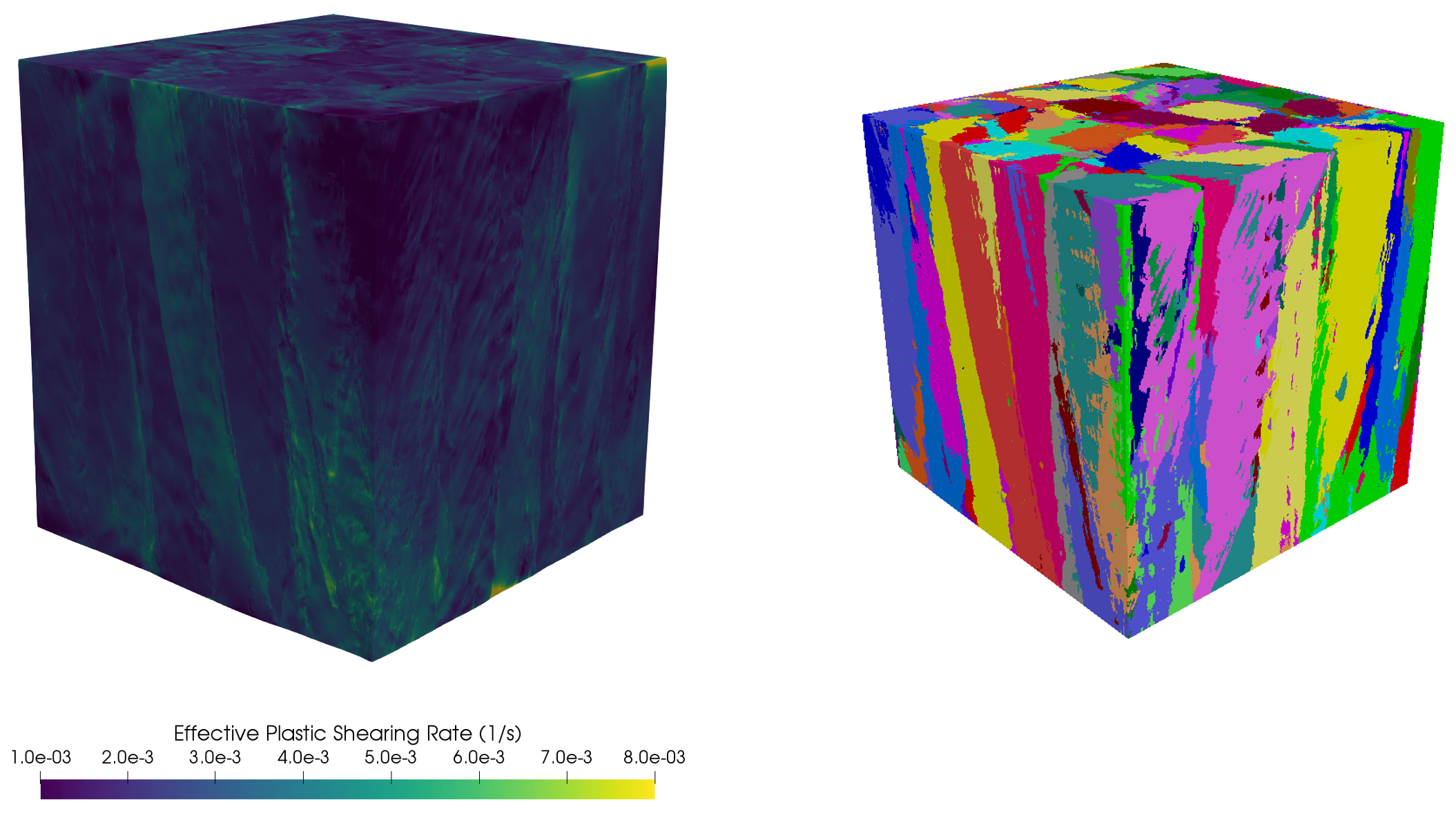}
\caption{A representative microstructure within an additive manufactured part
over a 500 micron volume cube and discretized into 27 million linear hexahedron
elements. The highly heterogeneous effective plastic shearing rate is plotted
along with the microstructure where each crystal is represented as a different color.}
\label{fig:exaam}
\end{figure}

ExaConstit \cite{exaconstit} is a general implicit quasi-static non-linear solid mechanics velocity-based finite element application built on the MFEM framework \cite{mfem}.
This code is being developed at LLNL for the ExaAM project in the ECP, with the goals of connecting local additive manufactured microstructures to local macroscopic properties by means of crystal plasticity finite element methods.
\reva{
As part of a larger workflow of the ExaAM workflow to simulate the additive manufacturing process, ExaConstit and its constitutive library, ExaCMech \cite{ecmech}, needed be refactored to run on the GPU.
This refactoring was required to run the hundreds to thousands of high-fidelity simulations on exascale hardware in a timely manner for the larger workflow.
ExaCMech was ported over to the GPU using RAJA \code{forall} loops wrapped around the entire large constitutive kernel.
Within ExaConstit, the dominant computational cost lies within the linearized system solve within a Newton-Raphson scheme.
Therefore,the primary focus had been transitioning from a traditional full assembly method over to partial and element assembly methods.
This transition required the physics/constitutive calculations to be completely separated from the assembly method and called in a separate setup phase.
The setup phase is now responsible for calculating the updated stress and material tangent stiffness matrix.
Afterwords, these values can be incorporated into any of the runtime selected linear assembly methods.
Initial partial and element assembly formulations are based on \cite{Gupta1972} and \cite{Gupta1983}, respectively.
In order to keep compute kernels backend agnostic, MFEM's \code{forall} abstraction macros, which make use of the RAJA backends, and memory management capabilities were leveraged within ExaConstit for a vast majority of the compute kernels. Within ExaConstit, a few reduction operations were also converted to RAJA reduction policies to take advantage of the GPU as these are not available within the MFEM API.
The end result of this refactoring is a $\sim$14.5x speed-up when using the GPU element assembly over the CPU full assembly on Summit.
For an ExaAM challenge problem sized linear hexahedron mesh of 6.7 million elements that undergoes 5\% monotonic strain, the GPU port and assembly improvements result in a runtime decrease of roughly 35 node-hours down to 2.5 node-hours on 8 nodes of Summit for just the ExaConstit stage.
The results of these simulations will then be used determine the properties being used in the part-scale simulation of the larger ExaAM workflow being run on exascale hardware.
Finally from a physics point of view, these improvements are also enabling the ExaAM team to study the highly complex deformation processes that occur within additively manufactured microstructures, such as those shown in Figure \ref{fig:exaam}, at an unprecedented level of fidelity of typical crystal plasticity methods.}

\section{Conclusions} \label{sec:conclusions}

In this paper we described the development of GPU-oriented algorithms
for high-order finite element discretizations in the ECP CEED
projects. We presented the current GPU capabilities of several CEED
components, including libCEED, MAGMA, MFEM, libParanumal and Nek,
which can now run efficiently on both NVIDIA and AMD GPUs. We also
discussed some of the challenges of porting to exascale GPU
architectures and presented application results that use the
CEED-developed GPU technologies.

\section*{Acknowledgments}

This research is supported by the Exascale Computing Project (17-SC-20-SC), a
collaborative effort of two U.S. Department of Energy organizations (Office of
Science and the National Nuclear Security Administration) responsible for the
planning and preparation of a capable exascale ecosystem, including software,
applications, hardware, advanced system engineering and early testbed platforms,
in support of the nation's exascale computing imperative.

The research used resources of the Argonne Leadership Computing Facility, which
is supported by the U.S. Department of Energy, Office of Science, under Contract
DE-AC02-06CH11357. This research also used resources of the Oak Ridge Leadership
Computing Facility at Oak Ridge National Laboratory, which is supported by the
Office of Science of the U.S. Department of Energy under Contract DE-AC05-00OR22725.
Work performed under the auspices of the U.S. Department of Energy under
Contract DE-AC52-07NA27344 (LLNL-JRNL-816034).

\bibliography{ceed}

\end{document}